\documentclass[journal]{IEEEtran}
\usepackage{cite}
\ifCLASSINFOpdf
  \usepackage[pdftex]{graphicx}
\else
  \usepackage[dvips]{graphicx}
\fi
\usepackage[draft]{changes}
\setremarkmarkup{\textsuperscript{#2}}
\newcommand{\stkout}[1]{\ifmmode\text{\sout{\ensuremath{#1}}}\else\sout{#1}\fi}
\setdeletedmarkup{\stkout{#1}}
\usepackage{amsmath}
\usepackage{algorithmic}
\usepackage{algorithm}
\usepackage{array}
\usepackage{amsfonts}
\usepackage{amssymb}
\usepackage{bm}
\usepackage{color}

\ifCLASSOPTIONcompsoc
  \usepackage[caption=false,font=normalsize,labelfont=sf,textfont=sf]{subfig}
\else
  \usepackage[caption=false,font=footnotesize]{subfig}
\fi
\usepackage{url}

\begin{document}

\title{CT Super-resolution GAN Constrained by the Identical, Residual, and Cycle Learning Ensemble (GAN-CIRCLE)}

\author{Chenyu You, Guang Li, Yi Zhang,~\IEEEmembership{Senior Member, IEEE}, Xiaoliu Zhang, Hongming Shan, Shenghong Ju,\\ Zhen Zhao, Zhuiyang Zhang, Wenxiang Cong, Michael W. Vannier,~\IEEEmembership{Member, IEEE},\\ Punam K. Saha,~\IEEEmembership{Senior Member, IEEE}, and Ge Wang*,~\IEEEmembership{Fellow, IEEE}
\thanks{Asterisk indicates corresponding author.}
\thanks{C. You is with Departments of Bioengineering and Electrical Engineering, Stanford University, Stanford, CA, 94305 USA (e-mail: uniycy@stanford.edu)}
\thanks{G. Li, H. Shan, W. Cong, and G. Wang* are with Department of Biomedical Engineering, Rensselaer Polytechnic Institute, Troy, NY, 12180 USA (e-mail: lig10@rpi.edu, shanh@rpi.edu, congw@rpi.edu, wangg6@rpi.edu)}
\thanks{Y. Zhang is with the College of Computer Science, Sichuan University, Chengdu, 610065 China (e-mail: yzhang@scu.edu.cn)}
\thanks{X. Zhang is with Department of Electrical and Computer Engineering, University of Iowa, Iowa City, IA, 52246 USA, (email: xiaoliu-zhang@uiowa.edu)}
\thanks{S. Ju, Z. Zhao are with Jiangsu Key Laboratory of Molecular and Functional Imaging, Department of Radiology, Zhongda Hospital, Medical School, Southeast University, Nanjing, 210009 China (e-mail: jsh0836@hotmail.com, zhaozhen8810@126.com)}
\thanks{Z. Zhang is with Department of Radiology, Wuxi No.2 People's Hospital, Wuxi, 214000 China (e-mail: zhangzhuiyang@163.com)}
\thanks{M. W. Vannier is with Department of Radiology, University of Chicago, Chicago, IL, 60637 USA}
\thanks{P. K. Saha is with Department of Electrical and Computer Engineering and Radiology, University of Iowa, Iowa City, IA, 52246 USA, (email: pksaha@engineering.uiowa.edu)}
}

\maketitle

\begin{abstract}
Computed tomography (CT) is widely used in screening, diagnosis, and image-guided therapy for both clinical and research purposes. Since CT involves ionizing radiation, an overarching thrust of related technical research is development of novel methods enabling ultrahigh quality imaging with fine structural details while reducing the X-ray radiation. In this paper, we present a semi-supervised deep learning approach to accurately recover high-resolution (HR) CT images from low-resolution (LR) counterparts. Specifically, with the generative adversarial network (GAN) as the building block, we enforce the cycle-consistency in terms of the Wasserstein distance to establish a nonlinear end-to-end mapping from noisy LR input images to denoised and deblurred HR outputs. We also include the joint constraints in the loss function to facilitate structural preservation. In this deep imaging process, we incorporate deep convolutional neural network (CNN), residual learning, and network in network techniques for feature extraction and restoration. In contrast to the current trend of increasing network depth and complexity to boost the CT imaging performance, which limit its real-world applications by imposing considerable computational and memory overheads, we apply a parallel $\pmb{1\times1}$ CNN to compress the output of the hidden layer and optimize the number of layers and the number of filters for each convolutional layer. Quantitative and qualitative evaluations demonstrate that our proposed model is accurate, efficient and robust for super-resolution (SR) image restoration from noisy LR input images. In particular, we validate our composite SR networks on three large-scale CT datasets, and obtain promising results as compared to the other state-of-the-art methods.
\end{abstract}


 \ifCLASSOPTIONpeerreview
 \begin{center} \bfseries EDICS Category: 3-BBND \end{center}
 \fi
%
\IEEEpeerreviewmaketitle

\section{Introduction}
\IEEEPARstart{X}{-ray} computed tomography (CT) is one of the most popular medical imaging methods for screening, diagnosis, and image-guided intervention~\cite{brenner2001estimated}. Potentially, high-resolution (HR) CT (HRCT) imaging may enhance the fidelity of radiomic features as well. Therefore, super-resolution (SR) methods in the CT field are receiving a major attention~\cite{park2003super,greenspan2008super}. The image resolution of a CT imaging system is constrained by x-ray focal spot size, detector element pitch, reconstruction algorithms, and other factors. While physiological and pathological units in the human body are on an order of 10 microns, the in-plane and through-plane resolution of clinical CT systems are on an order of submillimeter or 1 $mm$~\cite{greenspan2008super,schwarzband2005point}. Even though the modern CT imaging and visualization software can generate any small voxels, the intrinsic resolution is still far lower than what is ideal in important applications such as early tumor characterization and coronary artery analysis~\cite{hassan2011technical}. Consequently, how to produce HRCT images at a minimum radiation dose level is a holy grail of the CT field.

\begin{figure*}[!ht]
\begin{center}
\includegraphics[width=7.0in]{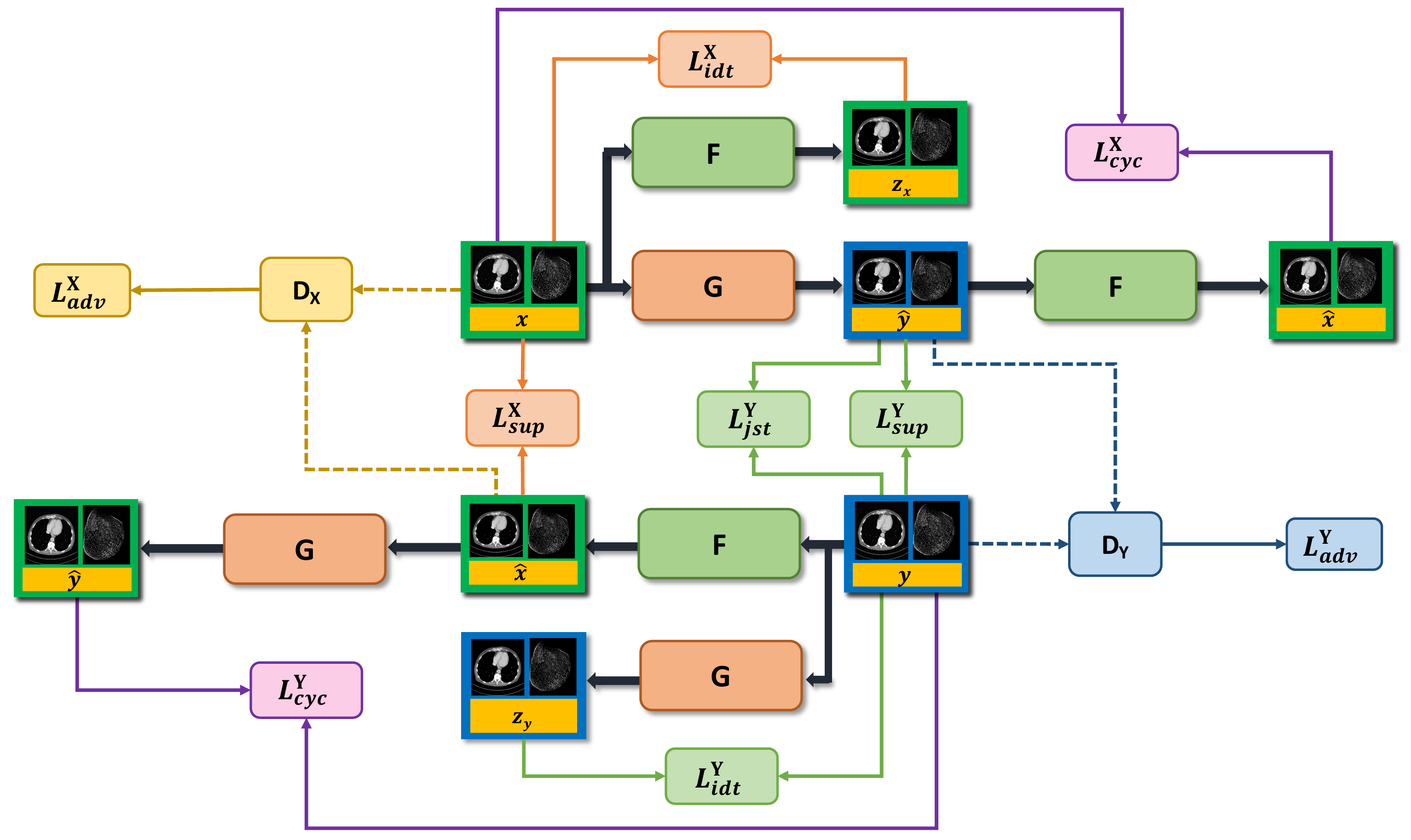}
\caption{Proposed GAN framework for SR CT imaging. Our approach uses two generators $G$ and $F$, and the corresponding adversarial discriminators $D_{X}$ and $D_{Y}$ respectively, where $X$ and $Y$ denote a LR CT image and $Y$ is the HR CT counterpart. To regularize the training and deblurring processes, we utilize the generator-adversarial loss ($adv$), cycle-consistency loss ($cyc$), identity loss ($idt$), and joint sparsifying transform loss ($jst$) synergistically. In the supervised/semi-supervised mode, we also apply a supervision loss ($sup$) on $G$ and $F$. For brevity, we denote $G: X\to Y$ and $F: Y \to X$ as $G$ and $F$ respectively.}\label{fig: overall}
\end{center}
\end{figure*}

In general, there are two strategies for improving CT image resolution: (1) hardware-oriented and (2) computational. First, more sophisticated hardware components can be used, including an x-ray tube with a fine focal spot size, detector elements of small pitch, and better mechanical precision for CT scanning. These hardware-oriented methods are generally expensive, increase the CT system cost and radiation dose, and compromise the imaging speed. Especially, it is well known that high X radiation dosage in a patient could induce genetic damages and cancerous diseases~\cite{brenner2007computed,de2004risk}. As a result, the second type of methods for resolution improvement~\cite{la2006penalized,vannier1996iterative,wang1998spiral,robertson1997total,jiang2003blind,jiang2002blind,wang2005blind} is more attractive, which is to obtain HRCT images from LRCT images. This computational deblurring job is a major challenge, representing a seriously ill-posed inverse problem~\cite{greenspan2008super,tian2011survey}. Our neural network approach proposed in this paper is computational, utilizing advanced network architectures. More details are as follows.

To reconstruct HRCT images, various algorithms were proposed. These algorithms can be broadly categorized into the following classes: (1)~\textit{Model-based reconstruction methods}~\cite{zhang2014model,bouman1996unified,yu2011fast,sauer1993local,thibault2007three}: These techniques explicitly model the image degradation process and regularize the reconstruction according to the characteristics of projection data. These algorithms promise an optimal image quality under the assumption that model-based priors can be effectively imposed; and (2)~\textit{Learning-based (before deep learning) SR methods}~\cite{yang2010image,wang2015learning,jiang2018super,zhang2012reconstruction,dong2011image}: These methods learn a nonlinear mapping from a training dataset consisting of paired LR and HR images to recover missing high-frequency details. Especially, sparse representation-based approaches have attracted an increasing interest since it exhibits strong robustness in preserving image features, suppressing noise and artifacts. Dong~\textit{et al.}~\cite{dong2011image} applied adaptive sparse domain selection and adaptive regularization to obtain excellent SR results in terms of both visual perceptions and PSNR. Zhang~\textit{et al.}~\cite{zhang2012reconstruction} proposed a patch-based technique for SR enhancements of 4D-CT images. These results demonstrate that learning-based SR methods can greatly enhance overall image quality but the outcomes may still lose image subtleties and yield blocky appearance.

Recently, deep learning (DL) has been instrumental for computer vision tasks~\cite{lecun1998gradient,lecun2015deep,krizhevsky2012imagenet}. Hierarchical features and representations derived from a convolutional neural network (CNN) are leveraged to enhance discriminative capacity of visual quality, thus people have started developing SR models for natural images~\cite{wang2017scalable,ledig2017photo,dong2016accelerating,LapSRN,shi2016real,Yuan_2018_CVPR_Workshops}. The key to the success of DL-based methods is its independence from explicit imaging models and backup by big domain-specific data. The image quality is optimized by learning features in an end-to-end manner. More importantly, once a CNN-based SR model is trained, achieving SR is a purely feed-forward propagation, which demands a very low computational overhead. 

In the medical imaging field, DL is an emerging approach which has exhibited a great potential~\cite{wang2017machine,wang2016perspective,wang2018image}. For several imaging modalities, DL-based SR methods were successfully  developed~\cite{chen2018efficient,yu2017computed,park2018computed,chaudhari2018super}. Chen~\textit{et al.}~\cite{chen2018efficient} proposed a deep densely connected super-resolution network to reconstruct HR brain magnetic resonance (MR) images. Chaudhari~\textit{et al.}~\cite{chaudhari2018super} developed a CNN-based network termed DeepResolve to learn a residual transformation from LR images to the corresponding HR images. More recently, Yu~\textit{et al.}~\cite{yu2017computed} proposed two advanced CNN-based models with a skip connection to promote high-frequency textures which are then fused with up-sampled images to produce SR images.

\begin{figure*}[!ht]
\begin{center}
\includegraphics[width=7.0in]{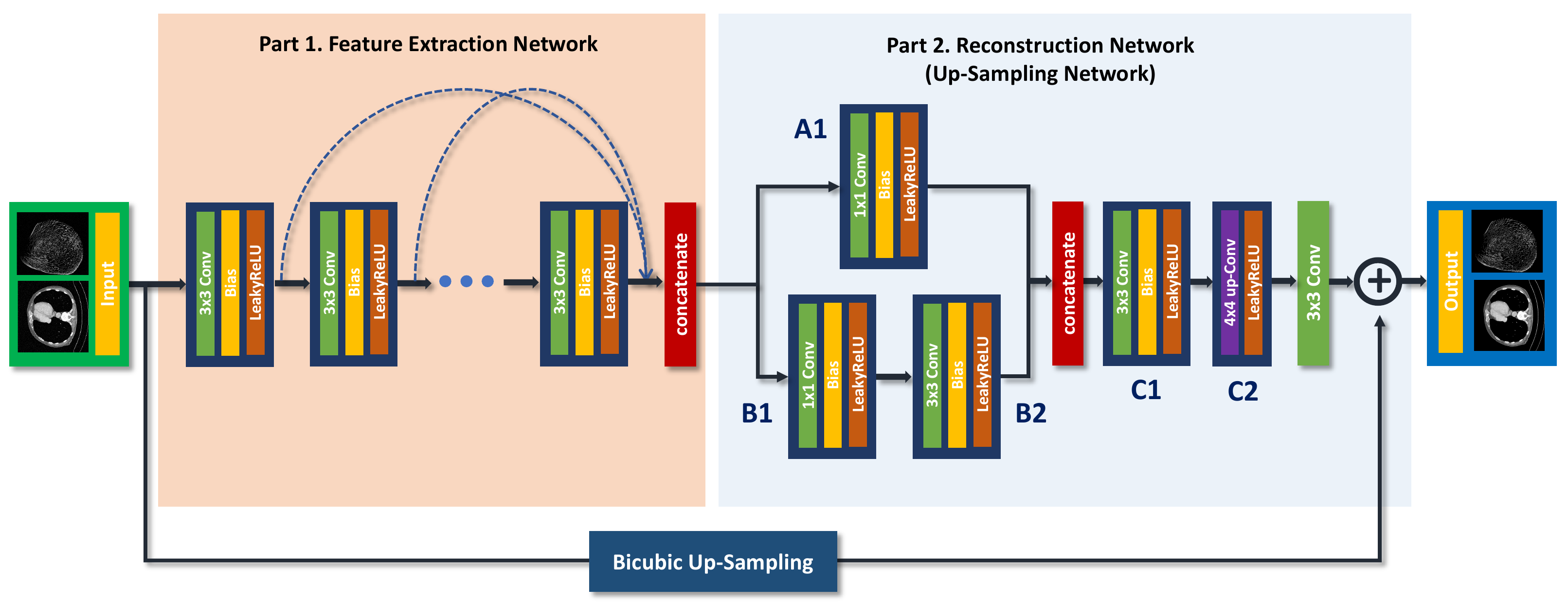}
\caption{Architecture of the SR generators. The generator is composed of feature extraction and reconstruction networks. The default stride is $1$, except for the $1^{st}$ feature blocks in which the stride for the conv layers is $2$. Up-scaling is performed to embed the residual layer for supervised training, and no interpolation method is used in the network for unsupervised feature learning.}
\label{fig: generator}
\end{center}
\end{figure*}

\begin{figure*}[!ht]
\begin{center}
\includegraphics[width=7.0in]{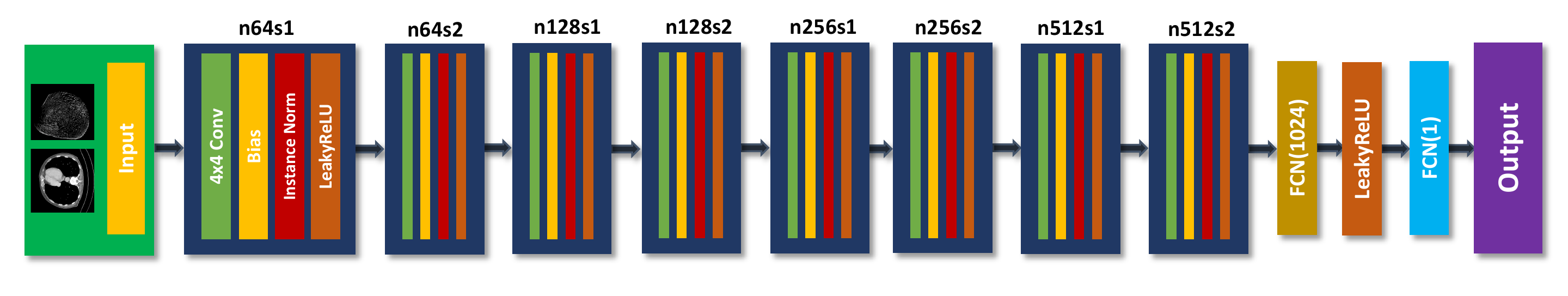}
\caption{A architecture of the discriminators. $n$ stands for the number of convolutional kernels, and $s$ stands for stride.~\textit{i.e.}, $n32s1$ means the convolutional layer of 32 kernels with stride 1.}
\label{fig: discriminator}
\end{center}
\end{figure*}

\begin{table*}[!ht]
\renewcommand{\arraystretch}{1.3}
\centering
\caption{Number of filters on each convolution (conv) layer of the generative network.}
\begin{tabular}{c c c c c c c c c c c c c c c c c c c} \\
\hline\hline
& \multicolumn{12}{c}{Feature extraction network} & \multicolumn{6}{c}{Reconstruction network} \\
& 1 & 2 &3 & 4 &5 &6 &7 &8 &9 &10 &11 &12 &A1 &B1 &B2 &C1 &C2 &Output \\
\cline{2-19}
$G$/$F$ & 64 & 54 & 48 & 43 & 39 & 35 & 31 & 28 & 25 & 22 & 18 & 16 & 24 & 8 & 8 & 32 & 16 & 1\\
\hline\hline
\end{tabular}
\label{table: generator_num_filter}
\end{table*}

Very recently, adversarial learning~\cite{goodfellow2014generative,goodfellow2016nips} has become increasingly popular, which enables the CNN to learn feature representations from complex data distributions, with unprecedented successes. Adversarial learning is performed based on a generative adversarial network (GAN), defined as a~\textit{mini-max} game in which two competing players are a generator $G$ and a discriminator $D$. In the game, $G$ is trained to learn a mapping from source images $\bm{x}$ in a source domain $X$ to target images $\bm{y}$ in the target domain $Y$. On the other hand, $D$ distinguishes the generated images $\hat{\bm{y}}$ and the target images $\bm{y}$ with a binary label. Once well trained, GAN is able to model a high-dimensional distribution of target images. Wolterink~\textit{et al.}~\cite{wolterink2017generative} proposed a unsupervised conditional GAN to optimize the nonlinear mapping from LR images to HR images, successfully enhancing the overall image quality. 

However, there are still several major limitations in the DL-based SR imaging. First, existing supervised DL-based algorithms cannot address blind SR tasks without LR-HR pairs. In clinical practice, the limited number of LR and HR CT image pairs makes the supervised learning methods impractical since it is infeasible to ask patients to take multiple CT scans with additional radiation doses for paired CT images. Thus, it is essential to resort to semi-supervised learning. Second, utilizing the adversarial strategy can push the generator to learn an inter-domain mapping and produce compelling target images ~\cite{nie2018medical} but there is a potential risk that the network may yield features that are not exhibited in target images due to the degeneracy of the mapping. Since the optimal $G$ is capable of translating $X$ to $\hat{Y}$ distributed identically to $\bm{y}$, the GAN network cannot ensure that the noisy input $\bm{x}$ and predicted output $\hat{\bm{y}}$ are paired in a meaningful way - there exist many mappings $G$ that may yield the same distribution over $\hat{Y}$. Consequently, the mapping is highly under-constrained. Furthermore, it is undesirable to optimize the adversarial objective in isolation: the model collapse problem may occur to map all inputs $\bm{x}$ to the same output image $\hat{\bm{y}}$~\cite{goodfellow2016nips,CycleGAN2017,kang2018cycle}. To address this issue, Cycle-consistent GANs (cycleGAN) was designed to improve the performance of generic GAN, and utilized for SR imaging~\cite{Yuan_2018_CVPR_Workshops}. Third, other limitations of GANs were also pointed out in~\cite{radford2015unsupervised,arjovsky2017wasserstein}. How to steer a GAN learning process is not easy since $G$ may collapse into a narrow distribution which cannot represent diverse samples from a real data distribution. Also, there is no interpretable metric for training progress. Fourth, as the number of layers increases, deep neural networks can derive a hierarchy of increasingly more complex and more abstract features. Frequently, to improve the SR imaging capability of a network, complex networks are often tried with hundreds of millions of parameters. However, given the associated computational overheads, they are hard to use in real-world applications. Fifth, local feature parts in the CT image have different scales. This feature hierarchy can provide more information to reconstruct images, but most DL-based methods~\cite{dong2016accelerating,LapSRN} neglect to use hierarchical features. Finally, the $L_{2}$ distance between $\hat{\bm{y}}$ and $\bm{y}$ is commonly used for the loss function to guide the training process of the network. However, the output optimized by the $L_{2}$ norm may suffer from over-smoothing  as discussed in~\cite{zhao2017loss,you2018structure}, since the $L_{2}$ distance means to maximizing the peak signal-to-noise rate (PSNR)~\cite{ledig2017photo}.

Motivated by the aforementioned drawbacks, in this study we made major efforts in the following aspects.~\textit{First}, we present a novel~\textbf{residual CNN-based network in the CycleGAN framework} to preserve high-resolution anatomical details with no task specific regularization. Specially, we utilize the cycle-consistency constraint to enforce a strong across-domain consistency between $X$ and $Y$.~\textit{Second}, to address the training problem of GANs~\cite{goodfellow2016nips,arjovsky2017wasserstein}, we use the Wasserstein distance or “Earth Moving” distance (EM distance) instead of the Jensen-Shannon (JS) divergence.~\textit{Third}, inspired by the recent work~\cite{yamanaka2017fast}, we optimize the network according to several fundamental designing principles to alleviate the computational overheads~\cite{he2016deep,huang2017densely,srivastava2014dropout}, which also helps prevent the network from over-fitting.~\textit{Fourth}, we cascade multiple layers to learn highly interpretable and disentangled hierarchical features. Moreover, we enable the information flow across the skip-connected layers to prevent gradient vanishing~\cite{he2016deep}.~\textit{Finally}, we employ the~$L_{1}$~norm instead of~$L_{2}$~norm to refine deblurring, and we propose to use a jointly constrained total variation-based regularization as well, which leverages the~\textit{prior} information to reduce the noise with a minimal loss in spatial resolution or anatomical information. Extensive experiments with three real datasets demonstrate that our proposed composite network can achieve an excellent CT SR imaging performance comparable to or better than that of the state-of-the-art methods~\cite{dong2016accelerating,ledig2017photo,shi2016real,LapSRN,yang2010image}.

\section{Methods}
\label{sec:methods}
Let us first review the SR problems in the medical imaging field. Then, we introduce the proposed adversarial nets framework and also present our SR imaging network architecture. Finally, we describe the optimization process.

\subsection{Problem Statement}
\label{subsec:problemsetting}
Let $\bm{x} \in X$ be an input LR image and a matrix $\bm{y} \in Y$ an output HR image, the conventional formulation of the ill-posed linear SR problem~\cite{yang2010image} can be formulated as
\begin{equation}
\bm{x} = \textit{S\,H}\,\bm{y} + \bm{\epsilon},
\end{equation}
where $\textit{S\,H}$ denote the down-sampling and blurring system matrix, and $\bm{\epsilon}$ the noise and other factors. Note that in practice, both the system matrix and not-modeled factors can be non-linear, instead of being linear (\textit{i.e.}, neither scalable nor additive).

Our goal is to computationally improve noisy LRCT images obtained under a low-dose CT (LDCT) protocol to HRCT images. The main challenges in recovering HRCT images can be listed as follows.~\textit{First}, LRCT images contain different or more complex spatial variations, correlations and statistical properties than natural images, which limit the SR imaging performance of the traditional methods.~\textit{Second}, the noise in raw projection data is introduced to the image domain during the reconstruction process, resulting in unique noise and artifact patterns. This creates difficulties for algorithms to produce the perfect image quality.~\textit{Finally}, since the sampling and degradation operations are coupled and ill-posed, SR tasks cannot be performed beyond a marginal degree using the traditional methods, which cannot effectively restore some fine features and suffer from the risk of producing blurry appearance and new artifacts. To address these limitations, here we develop an advanced neural network by composing a number of~\textbf{non-linear SR functional blocks} for SR CT (SRCT) imaging along with the~\textbf{residual module} to learn high-frequency details. Then, we perform~\textbf{adversarial learning in a cyclic manner} to generate perceptually and quantitatively superior SRCT images.

\subsection{Deep Cycle-Consistent Adversarial SRCT Model}
\subsubsection{\textbf{\textit{Cycle-Consistent Adversarial Model}}}
Current DL-based algorithms use feed-forward CNNs to learn non-linear mappings parametrized by $\pmb{\theta}$, which can be written as:
\begin{equation}
\hat{\bm{y}} = {G_{\bm{\theta}}}(\bm{x}) + \bm{\epsilon}.
\end{equation}
In order to obtain a decent $\bm{\hat{y}}$, a  suitable loss function must be specified to encourage $G_{\bm{\theta}}$ to generate a SR image based on the training samples so that \begin{equation}
\hat{\bm{\theta}} = \mathop{\arg\min}_{\bm{\theta}} \sum_{i} \mathcal{L}(\hat{\bm{y}}_{i},\bm{y}_{i}),
\end{equation}
where $(\bm{x}_{i},\bm{y}_i)$ are paired LRCT and HRCT images for training. To address the limitations mentioned in~\ref{subsec:problemsetting}, our Cyclic SRCT model is shown in Fig.~\ref{fig: overall}. The proposed model includes two generative mappings $G: X\to Y$ and $F: Y \to X$ given training samples $\bm{x}_{i} \in X$ and $\bm{y}_i \in Y$. Note that we denote the two mappings $G: X\to Y$ and $F: Y \to X$ as $G$ and $F$ respectively for brevity. The two mappings $G$ and $F$ are jointly trained to produce synthesized images in a way that confuse the adversarial discriminators $D_{Y}$ and $D_{X}$ respectively, which intend to identify whether the output of each generative mapping is real or artificial.~\textit{i.e.}, given an LRCT image $\bm{x}$, $G$ attempts to generate a synthesized image $\hat{\bm{y}}$ highly similar to a real image $\bm{y}$ so as to fool $D_{Y}$. In a similar way, $D_{X}$ attempts to discriminate between a reconstructed $\hat{\bm{x}}$ from $F$ and a real $\bm{x}$. The key idea is that the generators and discriminators are jointly/alternatively trained to improve their performance metrics synergistically. Thus, we have the following optimization problem:
\begin{equation}
\mathop{\min}_{G,F}\mathop{\max}_{D_{Y},D_{X}}\mathcal{L}_{\mathrm{GAN}}(G,D_{Y})+\mathcal{L}_{\mathrm{GAN}}(F,D_{X}).
\label{eq:obj_D_GAN}
\end{equation}
To enforce the mappings between the source and target domains and regularize the training procedure, our proposed network combines four types of loss functions:~\textbf{adversarial loss} (\textit{adv});~\textbf{cycle-consistency loss} (\textit{cyc});~\textbf{identity loss} (\textit{idt});~\textbf{joint sparsifying transform loss} (\textit{jst}).

\subsubsection{\textbf{Adversarial Loss}}
For~\textit{marginal matching}~\cite{goodfellow2014generative}, we employ adversarial losses to urge the generated images to obey the empirical distributions in the source and target domains. To improve the training quality, we apply the Wasserstein distance~\cite{gulrajani2017improved} instead of the negative log-likelihood used in~\cite{goodfellow2014generative}. Thus, we have the adversarial objective with respect to $G$:
\begin{multline}
\mathop{\min}_{G}\mathop{\max}_{D_{Y}}\mathcal{L}_{\mathrm{WGAN}}(D_{Y},G) = -\mathbb{E}_{\bm{y}}[D(\bm{y})]+\mathbb{E}_{\bm{x}}[D(G(\bm{x}))] \\ +\lambda \mathbb{E}_{\tilde{\bm{y}}}[(||\nabla_{\tilde{\bm{y}}}D(\tilde{\bm{y}})||_2-1)^2],
\label{eq: loss_DG_WGAN}
\end{multline}
where $\mathbb{E}(\cdot)$ denotes the expectation operator; the first two terms are in terms of the Wasserstein estimation, and the third term penalizes the deviation of the gradient norm of its input relative to one, $\tilde{\pmb{y}}$ is uniformly sampled along straight lines for pairs of $G(\bm{x})$ and $\bm{y}$, and $\lambda$ is a regularization parameter. A similar adversarial loss $\min_{F}\max_{D_{X}}\mathcal{L}_{\mathrm{WGAN}}(D_{X},F)$ is defined for marginal matching in the reverse direction.

\subsubsection{\textbf{Cycle Consistency Loss}}
Adversarial training is for marginal matching~\cite{goodfellow2014generative,goodfellow2016nips}. However, in these earlier studies~\cite{CycleGAN2017,Adda_CVPR2017}, it was found that using adversarial losses alone cannot ensure the learned function can transform a source input successfully to a target output. To promote the consistency between $F(G(\bm{x}))$ and $\bm{x}$, the cycle-consistency loss can be express as:
\begin{equation}
\begin{aligned}
\mathcal{L}_{\mathrm{CYC}}(G,F) &= \mathbb{E}_{\bm{x}} \left[ ||F\left(G(\bm{x})\right)-\bm{x}||_{1} \right] \\ &+ \mathbb{E}_{\bm{y}} \left[ ||G\left(F(\bm{y})\right)-\bm{y}||_{1} \right],
\label{eq: cyc_loss}
\end{aligned}
\end{equation}
where $||\cdot||_1$ denotes the $\mathcal{L}_{1}$ norm. Since the cycle consistency loss encourages $F(G(\bm{x})) \approx\bm{x}$ and $G(F(\bm{y}))\approx \bm{y}$, they are referred to as~\textit{forward cycle consistency} and~\textit{backward cycle consistency} respectively. The domain adaptation mapping refers to the~\textbf{cycle-reconstruction mapping}. In effect, it imposes shared-latent space constraints to encourage the source content to be preserved during the cycle-reconstruction mapping. In other words, the cycle consistency enforces latent codes deviating from the prior distribution in the cycle-reconstruction mapping. Additionally, the cycle consistency can help prevent the degeneracy in adversarial learning~\cite{li2017alice}. 

\subsubsection{\textbf{Identity Loss}}
Since a HR image should be a refined version of the LR counterpart, it is necessary to use the identity loss to regularize the training procedure~\cite{CycleGAN2017,kang2018cycle}. Compared with the $\mathcal{L}_{2}$ loss, the $\mathcal{L}_{1}$ loss does not over-penalize large differences or tolerate small errors between estimated and target images. Thus, the $\mathcal{L}_{1}$ loss is preferred to alleviate the limitations of the $\mathcal{L}_{2}$ loss in this context. Additionally, the $\mathcal{L}_{1}$ loss enjoys the same fast convergence speed as that of the $\mathcal{L}_{2}$ loss. The $\mathcal{L}_1$ loss is formulated as follows:
\begin{equation}
\begin{aligned}
\mathcal{L}_{\mathrm{IDT}}(G,F) &= \mathbb{E}_{\bm{y}} \left[ ||G\left(\bm{y}\right)-\bm{y}||_{1} \right] \\ &+ \mathbb{E}_{\bm{x}} \left[ ||\left(F(\bm{x}\right)-\bm{x}||_{1} \right].
\label{eq: loss_idt}
\end{aligned}
\end{equation}
We follow the same training baseline as in~\cite{kang2018cycle}; \textit{i.e.}, in the bi-directional mapping, the size of $G(\bm{y})$ (or $F(\bm{x})$) is the same as that of $\bm{y}$ (or $\bm{x}$).

\subsubsection{\textbf{Joint Sparsifying Transform Loss}}
The total variation (TV) has demonstrated the state-of-the-art performance in promoting image sparsity and reducing noise in piecewise-constant images~\cite{rudin1992nonlinear,sidky2008image,chen2008prior,vogel1996iterative,song2007sparseness,yang2010high,luo2018interior}. To express image sparsity, we formulate a nonlinear TV-based loss with the joint constraints as follows:
\begin{equation}
\begin{aligned}
\mathcal{L}_{\mathrm{JST}}(G) &= \tau\|G(\bm{x})\|_{\mathrm{TV}} + (1-\tau) \|\bm{y}-G(\bm{x})\|_{\mathrm{TV}},
\label{eq: tv_loss}
\end{aligned}
\end{equation}
where $\tau$ is a scaling factor. Intuitively, the above constrained minimization combines two components: the first component is used for sparsifying reconstructed images and alleviating conspicuous artifacts, and the second helps preserve anatomical characteristics by minimizing the difference image $\bm{y}-G(\bm{x})$. Essentially, these two components require a joint minimization under the bidirectional constraints. In this paper, the control parameter $\tau$ was set to $0.5$. In the case of $\tau=1$, the $\mathcal{L}_{\mathrm{JST}}(G)$ is regarded as the conventional $\mathrm{TV}$ loss.

\subsubsection{\textbf{Overall Objective Function}}
In the training process, our proposed network is fine-tuned in an end-to-end manner to minimize the following overall objective function:
\begin{equation}
\begin{aligned}
\mathcal{L}_{\mathrm{GAN-CIRCLE}} &= \mathcal{L}_{\mathrm{WGAN}}(D_{Y},G)+\mathcal{L}_{\mathrm{WGAN}}(D_{X},F) \\ &+\lambda_1 \mathcal{L}_{\mathrm{CYC}}(G,F)+\lambda_2 \mathcal{L}_{\mathrm{IDT}}(G,F)\\ &+\lambda_3 \mathcal{L}_{\mathrm{JST}}(G),
\label{eq: loss_final}
\end{aligned}
\end{equation}
where $\lambda_1$, $\lambda_2$, and $\lambda_3$ are parameters for balancing among different penalties. We call this modified cycleGAN as the GAN-CIRCLE as summarized in the title of this paper.

\subsubsection{\textbf{Supervised learning with GAN-CIRCLE}}
In the case where we have access to paired dataset, we can render SRCT problems to train our model in a supervised fashion. Given the training paired data from the true joint,~\textit{i.e.}, $(\bm{x},\bm{y})\sim P_{data}(\bm{X},\bm{Y})$, we can define a supervision loss as follows: 
\begin{equation}
\begin{aligned}
\mathcal{L}_{\mathrm{SUP}}(G,F) &= \mathbb{E}_{(\bm{x},\bm{y})} \left[ ||G\left(\bm{x}\right)-\bm{y}||_{1} \right] \\ &+ \mathbb{E}_{(\bm{x},\bm{y})} \left[ ||\left(F(\bm{y}\right)-\bm{x}||_{1} \right].
\label{eq: L_{sup}_loss}
\end{aligned}
\end{equation}

\begin{figure*}[!t]
\centering
\subfloat[Ground-truth HR]{\includegraphics[width=1.7in]{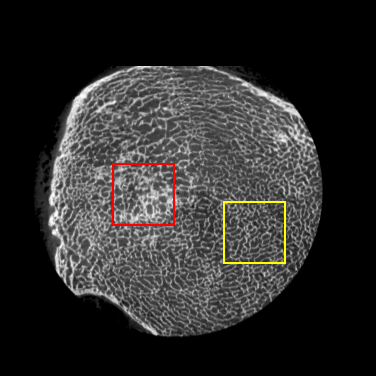}\label{fig: SR1}}
\subfloat[Noisy LR]{\includegraphics[width=1.7in]{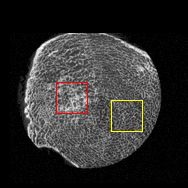}\label{fig: LR1}}
\subfloat[NN$^{\pmb{+}}$]{\includegraphics[width=1.7in]{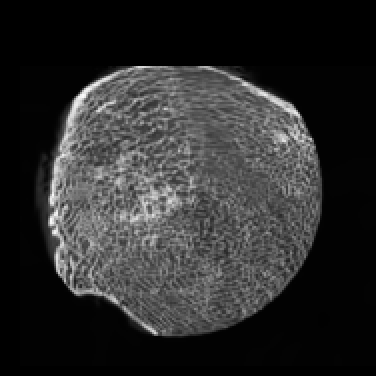}\label{fig: nn1}}
\subfloat[Bilinear$^{\pmb{+}}$]{\includegraphics[width=1.7in]{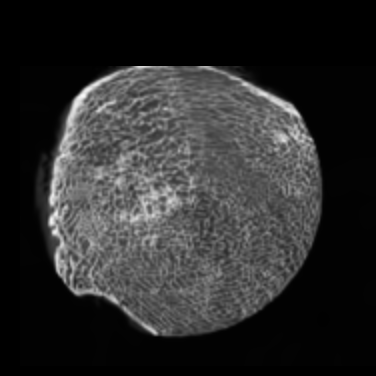}\label{fig: br1}}

\subfloat[Bicubic$^{\pmb{+}}$]{\includegraphics[width=1.7in]{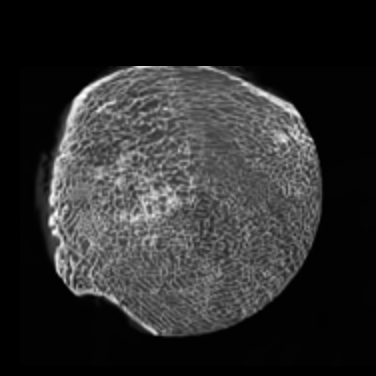}\label{fig: bc1}}
\subfloat[Lanczos$^{\pmb{+}}$]{\includegraphics[width=1.7in]{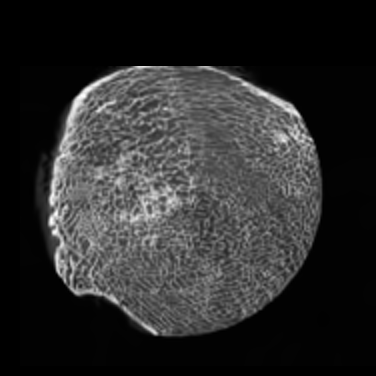}\label{fig: ls1}}
\subfloat[A$^{\pmb{+}}$]{\includegraphics[width=1.7in]{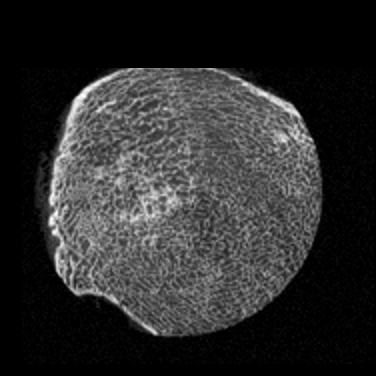}\label{fig: aplus}}
\subfloat[FSRCNN]{\includegraphics[width=1.7in]{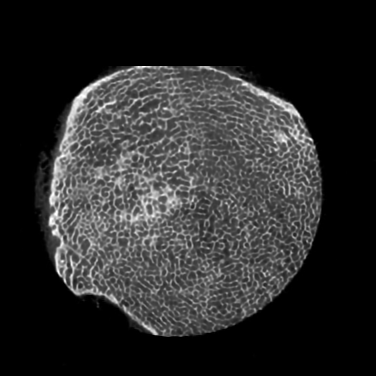}\label{fig: FSRCNN1}}

\subfloat[ESPCN]{\includegraphics[width=1.7in]{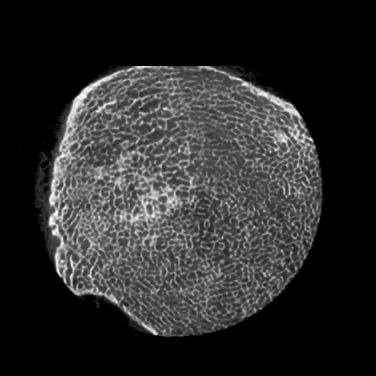}\label{fig: ESPCN1}}
\subfloat[LapSRN]{\includegraphics[width=1.7in]{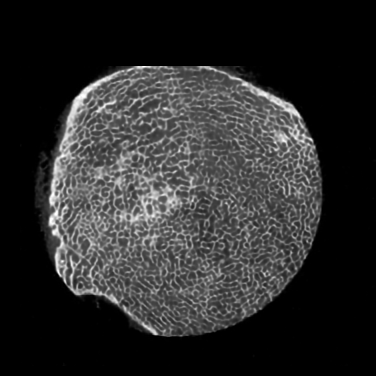}\label{fig: LapSRN1}}
\subfloat[SRGAN]{\includegraphics[width=1.7in]{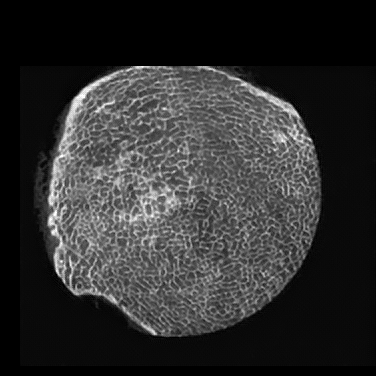}\label{fig: SRGAN1}}
\subfloat[G-Fwd]{\includegraphics[width=1.7in]{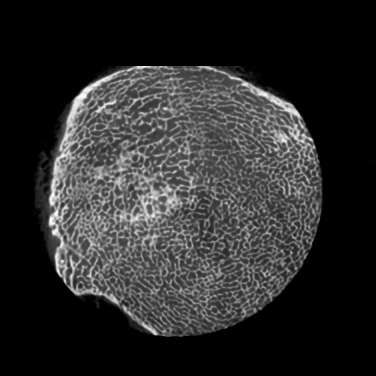}\label{fig: DCSCN1}}

\subfloat[G-Adv]{\includegraphics[width=1.7in]{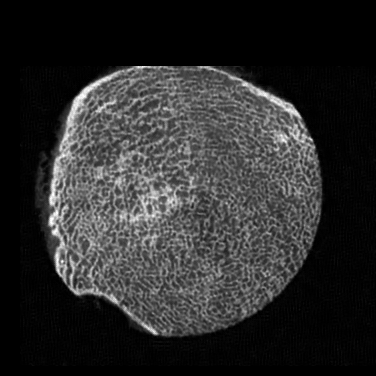}\label{fig: GAN_DCSCN1}}
\subfloat[GAN-CIRCLE]{\includegraphics[width=1.7in]{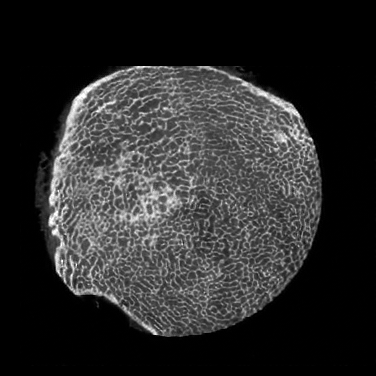}\label{fig: cWGAN_sup1}}
\subfloat[GAN-CIRCLE$^{s}$]{\includegraphics[width=1.7in]{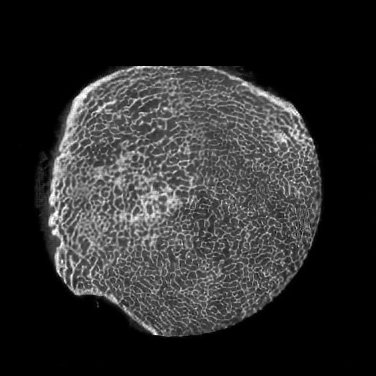}\label{fig: cWGAN_semisup1}}
\subfloat[GAN-CIRCLE$^{u}$]{\includegraphics[width=1.7in]{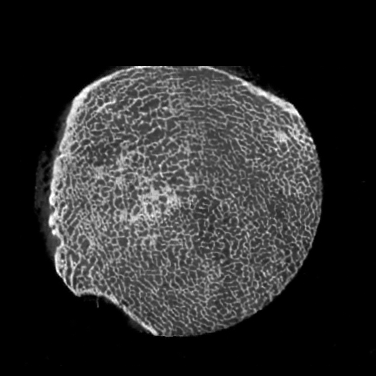}\label{fig: cWGAN_unsup1}}

\caption{Visual comparsion of SRCT Case~$1$ from the Tibia dataset. The restored bony structures are shown in the red and yellow boxes in Fig.~\ref{fig: example1_roi}.
The display window is [-900, 2000]~HU.}
\label{fig: example1}
\end{figure*}

\begin{figure*}[!t]
        \begin{minipage}[b]{.70\textwidth}
            \centering
            \subfloat[\scriptsize {Original HR}]{\includegraphics[width=0.80in]{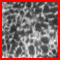}}~
            \subfloat[\scriptsize {Noisy LR}]{\includegraphics[width=0.80in]{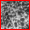}}~
            \subfloat[\scriptsize {NN$^{\pmb{+}}$}]{\includegraphics[width=0.80in]{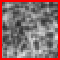}}~
            \subfloat[\scriptsize {Bilinear$^{\pmb{+}}$}]{\includegraphics[width=0.80in]{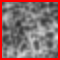}}~
            \subfloat[\scriptsize {Bicubic$^{\pmb{+}}$}]{\includegraphics[width=0.80in]{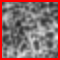}}~
            \subfloat[\scriptsize {Lanczos$^{\pmb{+}}$}]{\includegraphics[width=0.80in]{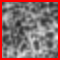}}~
            \subfloat[\scriptsize {A$^{\pmb{+}}$}]{\includegraphics[width=0.80in]{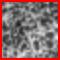}}~
            \subfloat[\scriptsize {FSRCNN}]{\includegraphics[width=0.80in]{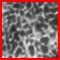}}\\[+.5ex]
            \subfloat[\scriptsize {ESPCN}]{\includegraphics[width=0.80in]{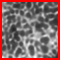}}~
            \subfloat[\scriptsize {LapSRN}]{\includegraphics[width=0.80in]{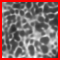}}~
            \subfloat[\scriptsize {SRGAN}]{\includegraphics[width=0.80in]{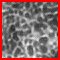}\label{fig:srgan1_roi1}}~
            \subfloat[\scriptsize {G-Fwd}]{\includegraphics[width=0.80in]{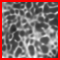}}~
            \subfloat[\scriptsize {G-Adv}]{\includegraphics[width=0.80in]{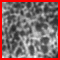}}~
            \subfloat[\scriptsize {\mbox{GAN-CIRCLE}}]{\includegraphics[width=0.80in]{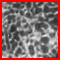}}~
            \subfloat[\scriptsize {\mbox{GAN-CIRCLE$^{s}$}}]{\includegraphics[width=0.80in]{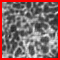}}~
            \subfloat[\scriptsize {\mbox{GAN-CIRCLE$^{u}$}}]{\includegraphics[width=0.80in]{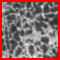}}
        \end{minipage}
        \label{fig: eg1_roi1}\\[0.5ex]
        \begin{minipage}[b]{.70\textwidth}
            \centering
            \subfloat[\scriptsize {Original HR}]{\includegraphics[width=0.80in]{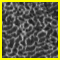}}~
            \subfloat[\scriptsize {Noisy LR}]{\includegraphics[width=0.80in]{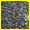}}~
            \subfloat[\scriptsize {NN$^{\pmb{+}}$}]{\includegraphics[width=0.80in]{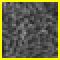}}~
            \subfloat[\scriptsize {Bilinear$^{\pmb{+}}$}]{\includegraphics[width=0.80in]{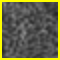}}~
            \subfloat[\scriptsize {Bicubic$^{\pmb{+}}$}]{\includegraphics[width=0.80in]{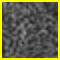}}~
            \subfloat[\scriptsize {Lanczos$^{\pmb{+}}$}]{\includegraphics[width=0.80in]{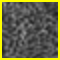}}~
            \subfloat[\scriptsize {A$^{\pmb{+}}$}]{\includegraphics[width=0.80in]{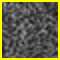}}~
            \subfloat[\scriptsize {FSRCNN}]{\includegraphics[width=0.80in]{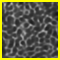}}\\[+.5ex]
            \subfloat[\scriptsize {ESPCN}]{\includegraphics[width=0.80in]{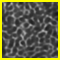}}~
            \subfloat[\scriptsize {LapSRN}]{\includegraphics[width=0.80in]{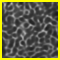}}~
            \subfloat[\scriptsize {SRGAN}]{\includegraphics[width=0.80in]{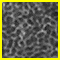}\label{fig:srgan1_roi2}}~
            \subfloat[\scriptsize {G-Fwd}]{\includegraphics[width=0.80in]{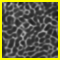}}~
            \subfloat[\scriptsize {G-Adv}]{\includegraphics[width=0.80in]{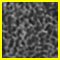}}~
            \subfloat[\scriptsize {\mbox{GAN-CIRCLE}}]{\includegraphics[width=0.80in]{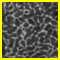}}~
            \subfloat[\scriptsize {\mbox{GAN-CIRCLE$^{s}$}}]{\includegraphics[width=0.80in]{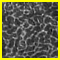}}~
            \subfloat[\scriptsize {\mbox{GAN-CIRCLE$^{u}$}}]{\includegraphics[width=0.80in]{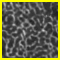}}
        \end{minipage}
        \label{fig: eg1_roi2}
\caption{Zoomed regions of interest~(ROIs) marked by the red rectangle in Fig.~\ref{fig: example1}. The restored image with GAN-CIRCLE reveals subtle structures better than the other variations of the proposed neural network, especially in the marked regions. The display window is [-900, 2000]~HU.}
\label{fig: example1_roi}
\end{figure*}

\subsection{Network Architecture}
\subsubsection{\textbf{Generative Networks}}
Although more layers and larger model size usually result in the performance gain, for real application we designed a lightweight model to validate the effectiveness of GAN-CIRCLE. The two generative networks $G$ and $F$ are shown in Fig.~\ref{fig: generator}. The network architecture has been optimized for SR CT imaging. It consists of two processing steams: the feature extraction network and the reconstruction network.

In the~\textbf{\textit{feature extraction network}}, we concatenate $12$ sets of non-linear SR feature blocks composed of $3\times 3$ Convolution (Conv) kernels, bias, Leaky ReLU, and a dropout layer. We utilize Leaky ReLU to prevent the `dead ReLU' problem thanks to the nature of leaky rectified linear units (Leaky ReLU): $\max(0,\bm{x})-\alpha \max(0,-\bm{x})$. Applying the dropout layer is to prevent overfitting. The number of filters are shown in Table~\ref{table: generator_num_filter}. In practice, we avoid normalization which is not suitable for SR, because we observe that it discards the range flexibility of the features. Then, to capture both local and the global image features, all outputs of the hidden layers are concatenated before the reconstruction network through skip connection. The skip connection helps prevent training saturation and overfitting. Diverse features which represent different details of the HRCT components can be constructed in the end of feature extraction network.

\begin{table*}[!t]
\renewcommand{\arraystretch}{1.0}
\centering
\caption{Quantitative evaluation of state-of-the-art SR algorithms. \textbf{{\color{red} Red}} and {\color{blue} \underline{blue}} indicate the best and the second best performance, respectively.}
\setlength{\tabcolsep}{5pt}
\begin{tabular}{c c c c c c c c c c c c c c c c}
\hline\hline
& \multicolumn{3}{c}{Tibia Case} && \multicolumn{3}{c}{Abdominal Case} && \multicolumn{3}{c}{Real Case~$1$} && \multicolumn{3}{c}{Real Case~$2$}\\
& PSNR & SSIM & IFC && PSNR & SSIM & IFC && PSNR & SSIM & IFC && PSNR & SSIM & IFC \\
\cline{2-4}\cline{6-8}\cline{10-12}\cline{14-16}
NN$^{\pmb{+}}$ & 24.754 & 0.645 & 2.785 && 26.566 & 0.592 & 1.919 && 28.072 & 0.798 & 0.246 && 27.903 & 0.798 & 0.234 \\
Bilinear$^{\pmb{+}}$ & 24.667 & 0.612 & 2.588 && 27.726 & 0.605 & 1.933 && 28.162 & 0.812 & 0.255 && 27.937 & 0.799 & 0.235 \\
Bicubic$^{\pmb{+}}$ & 25.641 & 0.662 & 2.835 && 28.599 & 0.619 & 2.008 && 28.117 & 0.805 & 0.250 && 27.929 & 0.798 & 0.235 \\
Lanczos$^{\pmb{+}}$ & 25.686 & 0.663 & 2.848 && 28.644 & 0.619 & 2.01 && 28.116 & 0.806 & 0.251 && 27.377 & 0.800 & 0.235  \\
A$^{\pmb{+}}$ & 26.496 & 0.696 & 3.028 && 28.154 & 0.589 & 1.899 && 27.877 & 0.804 & 0.249 && 27.037 & 0.778 & 0.236 \\
FSRCNN & 28.360 & {\color{blue}\underline{0.923}} & 3.533 && 30.950 & {\color{blue}\underline{0.924}} & 2.285 &&  \textbf{{\color{red}35.384}} & \textbf{{\color{red}0.830}} & 0.265 && 33.643 & \textbf{{\color{red}0.805}} & 0.237 \\
ESPCN & 28.361 & 0.921 & 3.520 && 30.507 & 0.923 & 2.252 && {\color{blue}\underline{35.378}} & \textbf{{\color{red}0.830}} & 0.278 && {\color{blue}\underline{33.689}} & \textbf{{\color{red}0.805}} & \textbf{{\color{red}0.245}} \\
LapSRN & {\color{blue}\underline{28.486}} & {\color{blue}\underline{0.923}} & 3.533 && {\color{blue}\underline{30.985}} &  \textbf{{\color{red}0.925}} & 2.299 && 35.372 & \textbf{{\color{red}0.830}} & 0.277 && \textbf{{\color{red}33.711}} & \textbf{{\color{red}0.805}} & {\color{blue}\underline{0.244}} \\
SRGAN & 21.924 & 0.389 & 1.620 && 28.550 & 0.871 & 1.925 && 33.002 & 0.737 & 0.232 && 31.775 & 0.701 & 0.220 \\
G-Fwd & \textbf{{\color{red}28.649}} & \textbf{{\color{red}0.931}} & 3.618 && \textbf{{\color{red}31.282}} &  \textbf{{\color{red}0.925}} & {\color{blue}\underline{2.348}} && 35.227 & {\color{blue}\underline{0.829}} & 0.276 && 33.589 & 0.803 & 0.236 \\
G-Adv & 26.945 & 0.676 & 2.999 && 26.930 & 0.889 & 1.765 && 32.518 & 0.725 & 0.199 && 31.712 & 0.700 & 0.210 \\
GAN-CIRCLE & 27.742 & 0.895 &  \textbf{{\color{red}3.944}} && 30.720 & {\color{blue}\underline{0.924}} &  \textbf{{\color{red}2.435}} && - & - & - && - & - & - \\
GAN-CIRCLE$^{s}$ & 27.071 & 0.887 & {\color{blue}\underline{3.893}} && 29.988 & 0.902 &  2.367 && 33.194 & {\color{blue}\underline{0.829}} & \textbf{{\color{red}0.285}} && 31.252 & {\color{blue}\underline{0.804}} & \textbf{{\color{red}0.245}} \\
GAN-CIRCLE$^{u}$ & 27.255 & 0.891 & 2.713 && 28.439 & 0.894 & 2.019 && 32.138 & 0.824 & {\color{blue}\underline{0.283}} && 30.641 & 0.796 & 0.232 \\
\hline\hline
\end{tabular}
\label{table: psnr&ssim&ifc}
\end{table*}

In the~\textbf{\textit{reconstruction network}}, we stack two reconstruction branches and integrate the information flows. Because all the outputs from the feature extraction network are densely connected, we propose a parallelized CNNs (Network in Network)~\cite{lin2013network} which utilize shallow multilayer perceptron (MLP) to perform a nonlinear projection in the spatial domain. There are several benefits with the Network in Network strategy. First, the $1\times 1$ Conv layer can significantly reduce the dimensionality of the filter space for faster computation with less information loss~\cite{lin2013network}. Second, the $1\times 1$ Conv layer can increase non-linearity of the network to learn a complex mapping better at the finer levels. For up-sampling, we adopt the transposed convolutional (up-sampling) layers~\cite{zeiler2010deconvolutional} by a scale of $2$. The last Conv layer fuses all the feature maps, resulting in an entire residual image containing mostly high-frequency details. In the supervised setting, the up-sampled image by the bicubic interpolation layer is combined (via element-wise addition) with the residual image to produce a HR output. In the unsupervised and semi-supervised settings, no interpolation is involved across the skip connection.

It should be noted that the generator $F$ shares the same architecture as $G$ in both the supervised and unsupervised scenarios. The default stride size is $1$. However, for unsupervised feature learning, the stride of the Conv layers is $2$  in the $1^{st}$ feature blocks. Also, for supervised feature learning, the stride of the Conv layers is $2$ in the $1^{st}$ and $2^{nd}$ feature blocks of $F$. We refer to the forward generative network $G$ as G-Forward.

\begin{figure*}[!t]
        \begin{minipage}[b]{0.25\textwidth}
            \centering
            \subfloat[\scriptsize{Original HR}]{\includegraphics[width=1.63in]{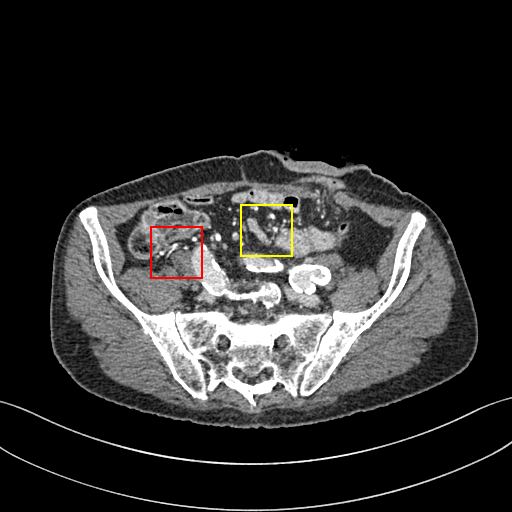}}  ~~
            \subfloat[\scriptsize{Noisy LR}]{\includegraphics[width=1.63in]{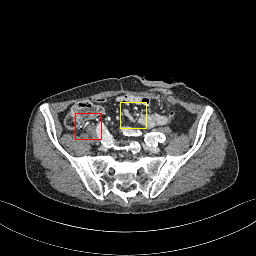}}  ~~
            \subfloat[\scriptsize{GAN-CIRCLE}]{\includegraphics[width=1.63in]{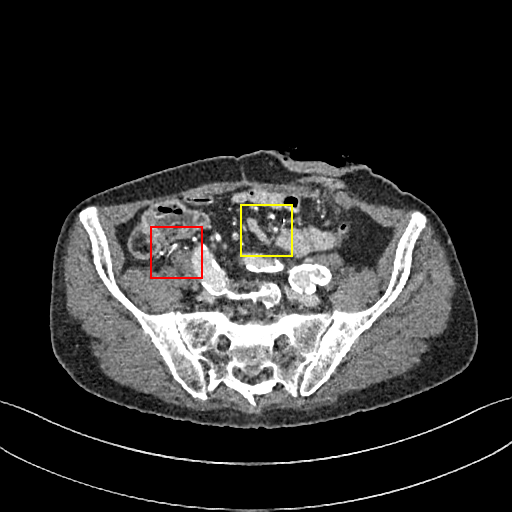}}  ~~
            \subfloat[\scriptsize{GAN-CIRCLE$^{s}$}]{\includegraphics[width=1.63in]{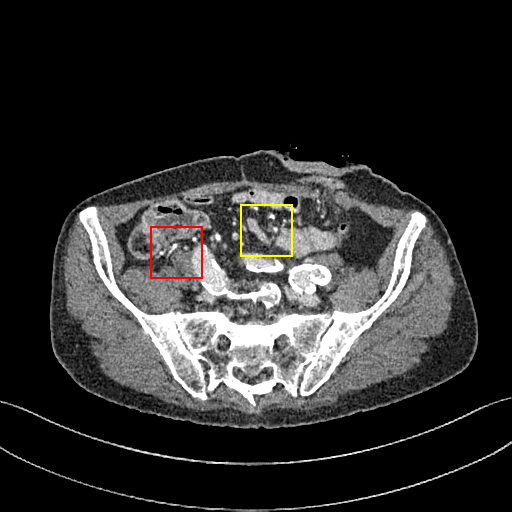}}
         \end{minipage}
         \label{fig: sr_lr2}\\[0.5ex]
        \begin{minipage}[b]{.70\textwidth}
            \centering
            \subfloat[\scriptsize {Original HR}]{\includegraphics[width=0.80in]{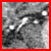}}~
            \subfloat[\scriptsize {Noisy LR}]{\includegraphics[width=0.80in]{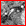}}~
            \subfloat[\scriptsize {NN$^{\pmb{+}}$}]{\includegraphics[width=0.80in]{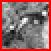}}~
            \subfloat[\scriptsize {Bilinear$^{\pmb{+}}$}]{\includegraphics[width=0.80in]{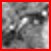}}~
            \subfloat[\scriptsize {Bicubic$^{\pmb{+}}$}]{\includegraphics[width=0.80in]{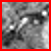}}~
            \subfloat[\scriptsize {Lanczos$^{\pmb{+}}$}]{\includegraphics[width=0.80in]{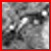}}~
            \subfloat[\scriptsize {A$^{\pmb{+}}$}]{\includegraphics[width=0.80in]{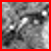}}~
            \subfloat[\scriptsize {FSRCNN}]{\includegraphics[width=0.80in]{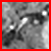}}\\[0.5ex]
            \subfloat[\scriptsize {ESPCN}]{\includegraphics[width=0.80in]{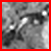}}~
            \subfloat[\scriptsize {LapSRN}]{\includegraphics[width=0.80in]{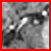}}~
            \subfloat[\scriptsize {SRGAN}]{\includegraphics[width=0.80in]{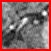}\label{fig:srgan2_roi1}}~
            \subfloat[\scriptsize {G-Fwd}]{\includegraphics[width=0.80in]{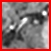}}~
            \subfloat[\scriptsize {G-Adv}]{\includegraphics[width=0.80in]{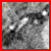}}~
            \subfloat[\scriptsize {\mbox{GAN-CIRCLE}}]{\includegraphics[width=0.80in]{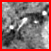}}~
            \subfloat[\scriptsize {\mbox{GAN-CIRCLE$^{s}$}}]{\includegraphics[width=0.80in]{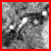}}~
            \subfloat[\scriptsize {\mbox{GAN-CIRCLE$^{u}$}}]{\includegraphics[width=0.80in]{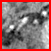}}
        \end{minipage}
        \label{fig: eg2_roi1}\\[0.5ex]
        \begin{minipage}[b]{.70\textwidth}
            \centering
            \subfloat[\scriptsize {Original HR}]{\includegraphics[width=0.80in]{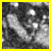}}~
            \subfloat[\scriptsize {Noisy LR}]{\includegraphics[width=0.80in]{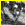}}~
            \subfloat[\scriptsize {NN$^{\pmb{+}}$}]{\includegraphics[width=0.80in]{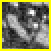}}~
            \subfloat[\scriptsize {Bilinear$^{\pmb{+}}$}]{\includegraphics[width=0.80in]{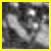}}~
            \subfloat[\scriptsize {Bicubic$^{\pmb{+}}$}]{\includegraphics[width=0.80in]{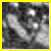}}~
            \subfloat[\scriptsize {Lanczos$^{\pmb{+}}$}]{\includegraphics[width=0.80in]{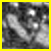}}~
            \subfloat[\scriptsize {A$^{\pmb{+}}$}]{\includegraphics[width=0.80in]{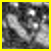}}~
            \subfloat[\scriptsize {FSRCNN}]{\includegraphics[width=0.80in]{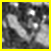}}\\[0.5ex]
            \subfloat[\scriptsize {ESPCN}]{\includegraphics[width=0.80in]{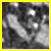}}~
            \subfloat[\scriptsize {LapSRN}]{\includegraphics[width=0.80in]{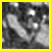}}~
            \subfloat[\scriptsize {SRGAN}]{\includegraphics[width=0.80in]{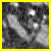}\label{fig:srgan2_roi2}}~
            \subfloat[\scriptsize {G-Fwd}]{\includegraphics[width=0.80in]{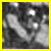}}~
            \subfloat[\scriptsize {G-Adv}]{\includegraphics[width=0.80in]{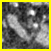}}~
            \subfloat[\scriptsize {\mbox{GAN-CIRCLE}}]{\includegraphics[width=0.80in]{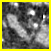}}~
            \subfloat[\scriptsize {\mbox{GAN-CIRCLE$^{s}$}}]{\includegraphics[width=0.80in]{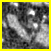}}~
            \subfloat[\scriptsize {\mbox{GAN-CIRCLE$^{u}$}}]{\includegraphics[width=0.80in]{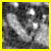}}
        \end{minipage}
        \label{fig: eg2_roi2}
\caption{Visual comparison of SRCT Case~$2$ from the abdominal dataset. The display window is [-160, 240]~HU. The restored anatomical features are shown in the red and yellow boxes. (\textbf{Zoomed for visual clarity}).}
\label{fig: example2}
\end{figure*}

\subsubsection{\textbf{Discriminative Networks}}
As shown in Fig.~\ref{fig: discriminator}, in reference to the recent successes with GANs~\cite{simonyan2014very,ledig2017photo}, $D$ is designed to have $4$ stages of Conv, bias, instance norm~\cite{instanceNorm2016} (IN) and Leaky ReLU, followed by two fully-connected layers, of which the first has $1024$ units and the other has a single output. In addition, inspired by~\cite{arjovsky2017wasserstein} no sigmoid cross entropy layer is applied by the end of $D$. We apply $4\times4$ filter size for the Conv layers which had different numbers of filters, which are $64, 64, 128, 128, 256, 256, 512, 512$ respectively.

\begin{figure*}[!t]
        \begin{minipage}[b]{0.25\textwidth}
            \centering
            \subfloat[\scriptsize{Original HR}]{\includegraphics[width=1.63in]{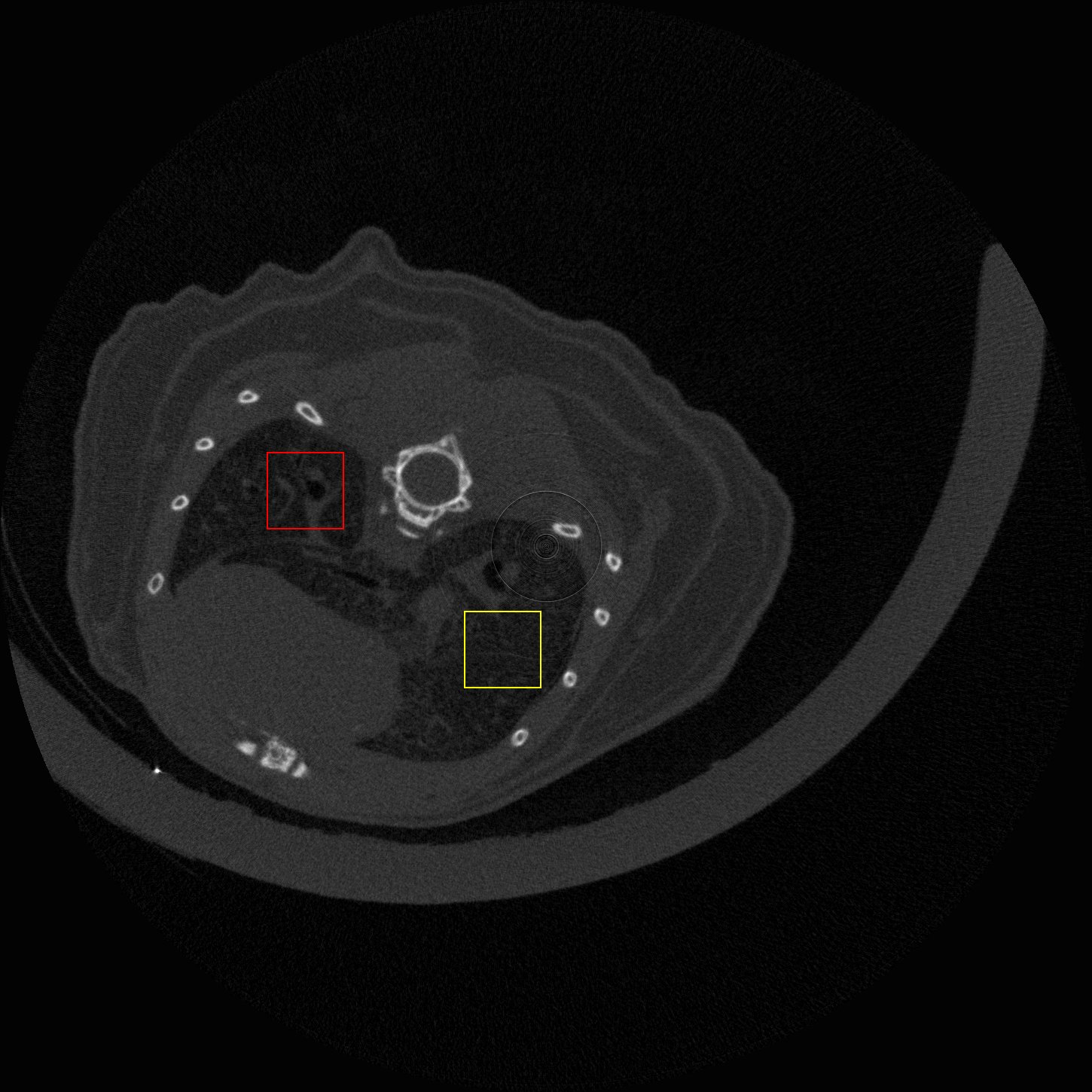}}  ~~
            \subfloat[\scriptsize{Noisy LR}]{\includegraphics[width=1.63in]{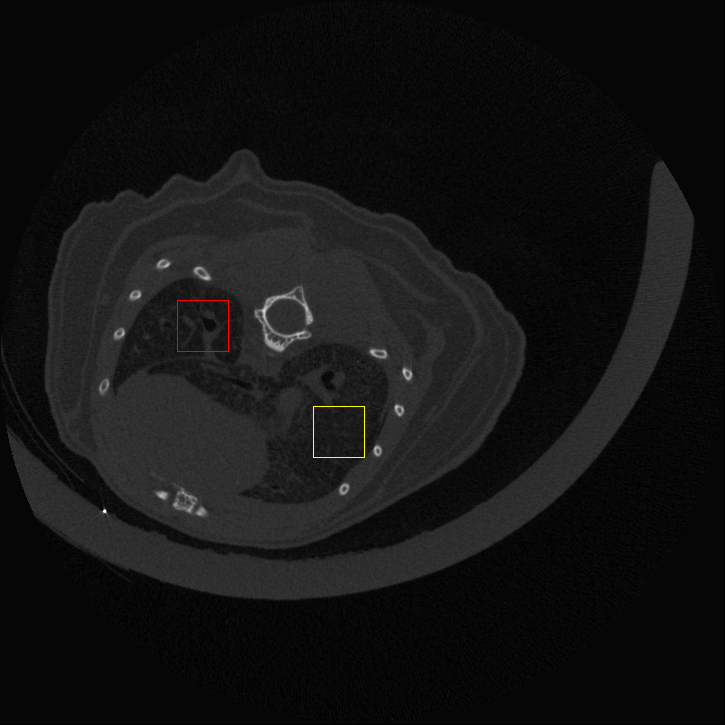}} ~~
            \subfloat[\scriptsize{GAN-CIRCLE$^{s}$}]{\includegraphics[width=1.63in]{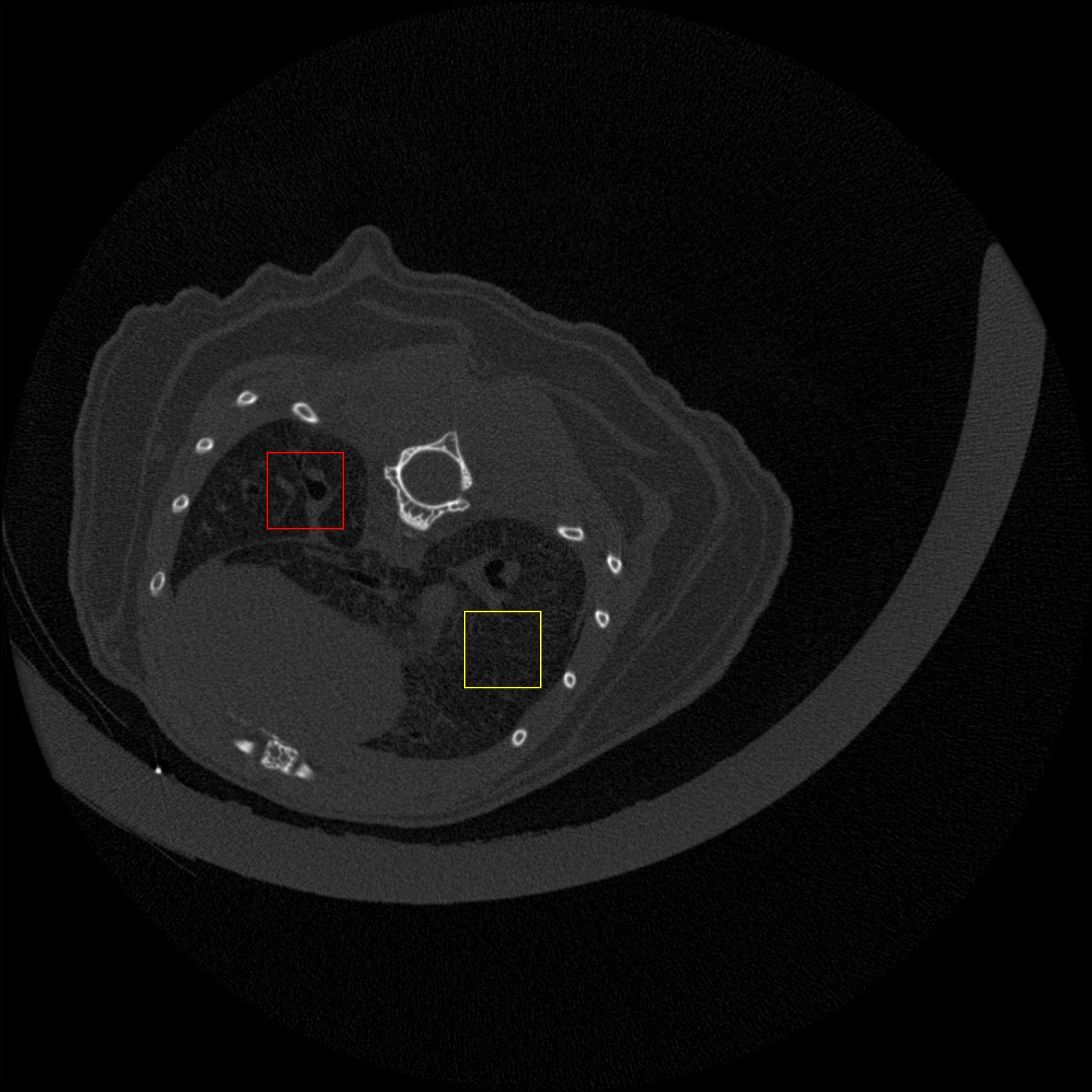}}  ~~
            \subfloat[\scriptsize{GAN-CIRCLE$^{u}$}]{\includegraphics[width=1.63in]{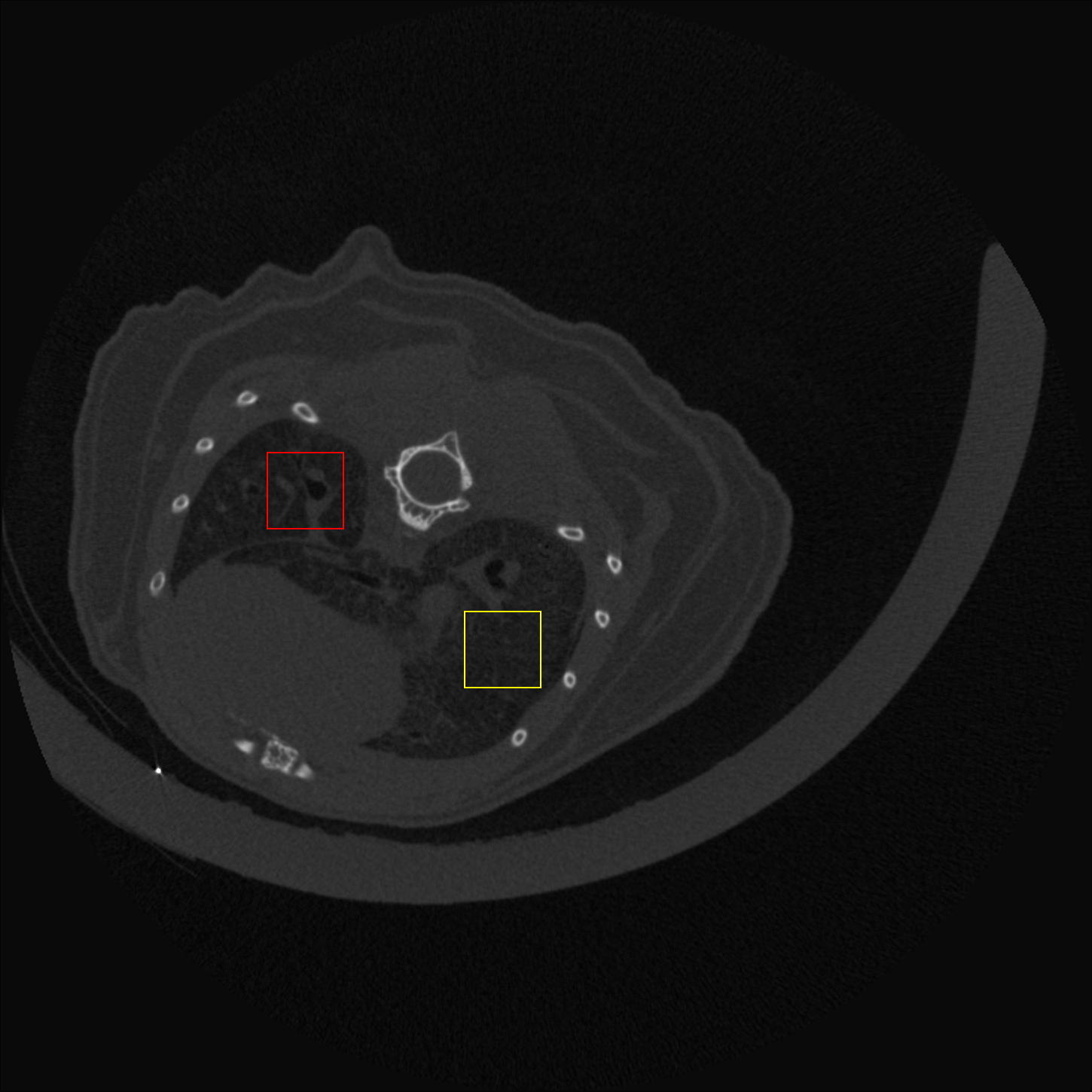}}
         \end{minipage}
         \label{fig: sr_lr5}\\[0.5ex]
        \begin{minipage}[b]{.80\textwidth}
            \centering
            \subfloat[\scriptsize {Noisy LR}]{\includegraphics[width=0.92in]{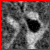}}~
            \subfloat[\scriptsize {NN$^{\pmb{+}}$}]{\includegraphics[width=0.92in]{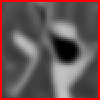}}~
            \subfloat[\scriptsize {Bilinear$^{\pmb{+}}$}]{\includegraphics[width=0.92in]{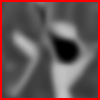}}~
            \subfloat[\scriptsize {Bicubic$^{\pmb{+}}$}]{\includegraphics[width=0.92in]{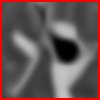}}~
            \subfloat[\scriptsize {Lanczos$^{\pmb{+}}$}]{\includegraphics[width=0.92in]{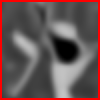}}~
            \subfloat[\scriptsize {A$^{\pmb{+}}$}]{\includegraphics[width=0.92in]{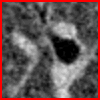}}~
            \subfloat[\scriptsize {FSRCNN}]{\includegraphics[width=0.92in]{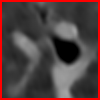}}\\[0.5ex]
            \subfloat[\scriptsize {ESPCN}]{\includegraphics[width=0.92in]{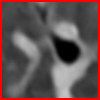}}~
            \subfloat[\scriptsize {LapSRN}]{\includegraphics[width=0.92in]{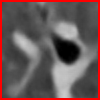}}~
            \subfloat[\scriptsize {SRGAN}]{\includegraphics[width=0.92in]{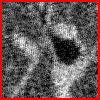}}~
            \subfloat[\scriptsize {G-Fwd}]{\includegraphics[width=0.92in]{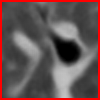}}~
            \subfloat[\scriptsize {G-Adv}]{\includegraphics[width=0.92in]{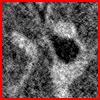}}~
            \subfloat[\scriptsize {\mbox{GAN-CIRCLE$^{s}$}}]{\includegraphics[width=0.92in]{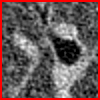}}~
            \subfloat[\scriptsize {\mbox{GAN-CIRCLE$^{u}$}}]{\includegraphics[width=0.92in]{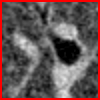}}
        \end{minipage}
        \label{fig: eg5_roi1}\\[0.5ex]
        \begin{minipage}[b]{.80\textwidth}
            \centering
            \subfloat[\scriptsize {Noisy LR}]{\includegraphics[width=0.92in]{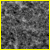}}~
            \subfloat[\scriptsize {NN$^{\pmb{+}}$}]{\includegraphics[width=0.92in]{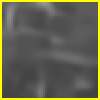}}~
            \subfloat[\scriptsize {Bilinear$^{\pmb{+}}$}]{\includegraphics[width=0.92in]{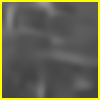}}~
            \subfloat[\scriptsize {Bicubic$^{\pmb{+}}$}]{\includegraphics[width=0.92in]{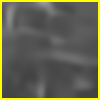}}~
            \subfloat[\scriptsize {Lanczos$^{\pmb{+}}$}]{\includegraphics[width=0.92in]{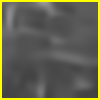}}~
            \subfloat[\scriptsize {A$^{\pmb{+}}$}]{\includegraphics[width=0.92in]{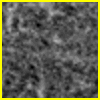}}~
            \subfloat[\scriptsize {FSRCNN}]{\includegraphics[width=0.92in]{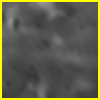}}\\[0.5ex]
            \subfloat[\scriptsize {ESPCN}]{\includegraphics[width=0.92in]{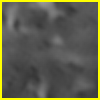}}~
            \subfloat[\scriptsize {LapSRN}]{\includegraphics[width=0.92in]{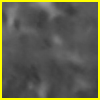}}~
            \subfloat[\scriptsize {SRGAN}]{\includegraphics[width=0.92in]{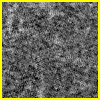}}~
            \subfloat[\scriptsize {G-Fwd}]{\includegraphics[width=0.92in]{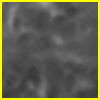}}~
            \subfloat[\scriptsize {G-Adv}]{\includegraphics[width=0.92in]{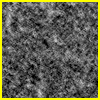}}~
            \subfloat[\scriptsize {\mbox{GAN-CIRCLE$^{s}$}}]{\includegraphics[width=0.92in]{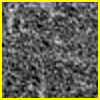}}~
            \subfloat[\scriptsize {\mbox{GAN-CIRCLE$^{u}$}}]{\includegraphics[width=0.92in]{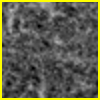}}
        \end{minipage}
        \label{fig: eg5_roi2}
\caption{Visual comparison of SRCT Case~$3$ from the real dataset. The display window is [180, 4096]~HU. The restored anatomical features are shown in the red and yellow boxes. (\textbf{Zoomed for visual clarity}).}
\label{fig: example5}
\end{figure*}

\section{Experiments and results}
\label{sec:exp}
We discuss our experiments in this section. We first introduce the datasets we utilize and then describe the implementation details and parameter settings in our proposed methods. We also compare our proposed algorithms with the state-of-the-art SR methods~\cite{shi2016real,LapSRN,ledig2017photo,dong2016accelerating} quantitatively and qualitatively. We further evaluate our results in reference to the state-of-the-art, and demonstrate the robustness of our methods in the real SR scenarios. Finally, we present the detailed diagnostic quality assessments from expert radiologists. Note that we use the default parameters of all the evaluated methods.

\subsection{Training Datasets}
\label{subsec:dataset}
In this study, we used two high-quality sets of training images to demonstrate the fidelity and robustness of the proposed GAN-CIRCLE. As shown in Figs.~\ref{fig: example1}~-~\ref{fig: example2}, these two datasets are of very different characteristics.

\subsubsection{\textbf{Tibia dataset}}
This micro-CT image dataset reflects twenty-five fresh-frozen cadaveric ankle specimens which were removed at mid-tibia from 17 body donors (mean age at death~$\pm$~SD: $79.6\,\pm\,13.2$ Y; $9$ female). After the soft tissue were removed and the tibia was dislocated from the ankle joint, each specimen was scanned on a Siemens microCAT II (Preclinical Solutions, Knoxville, TN, USA) in the cone beam imaging geometry. The micro-CT parameters are briefly summarized as follows: a tube voltage $100$ kV, a tube current $200$ mAs, $720$ projections over a range of $220$ degrees, an exposure time of $1.0$ sec per projection, and the filter backprojection (FBP) method was utilized to produce $28.8~\mu m$ isotropic voxels. Since CT images are not isotropic in each direction, for convenience of our previous analysis~\cite{chen2018quantitative}, we convert micro-CT images to $150~\mu m$ using a windowed sync interpolation method. In this study, the micro-CT images we utilized as HR images were prepared at $150~\mu m$ voxel size, as the target for SR imaging based of the corresponding LR images at $300~\mu m$ voxel size. The full description is in~\cite{chen2018quantitative}. We target $1$X resolution improvement.  

\begin{figure*}[!t]
        \begin{minipage}[b]{0.25\textwidth}
            \centering
            \subfloat[\scriptsize{Original HR}]{\includegraphics[width=1.63in]{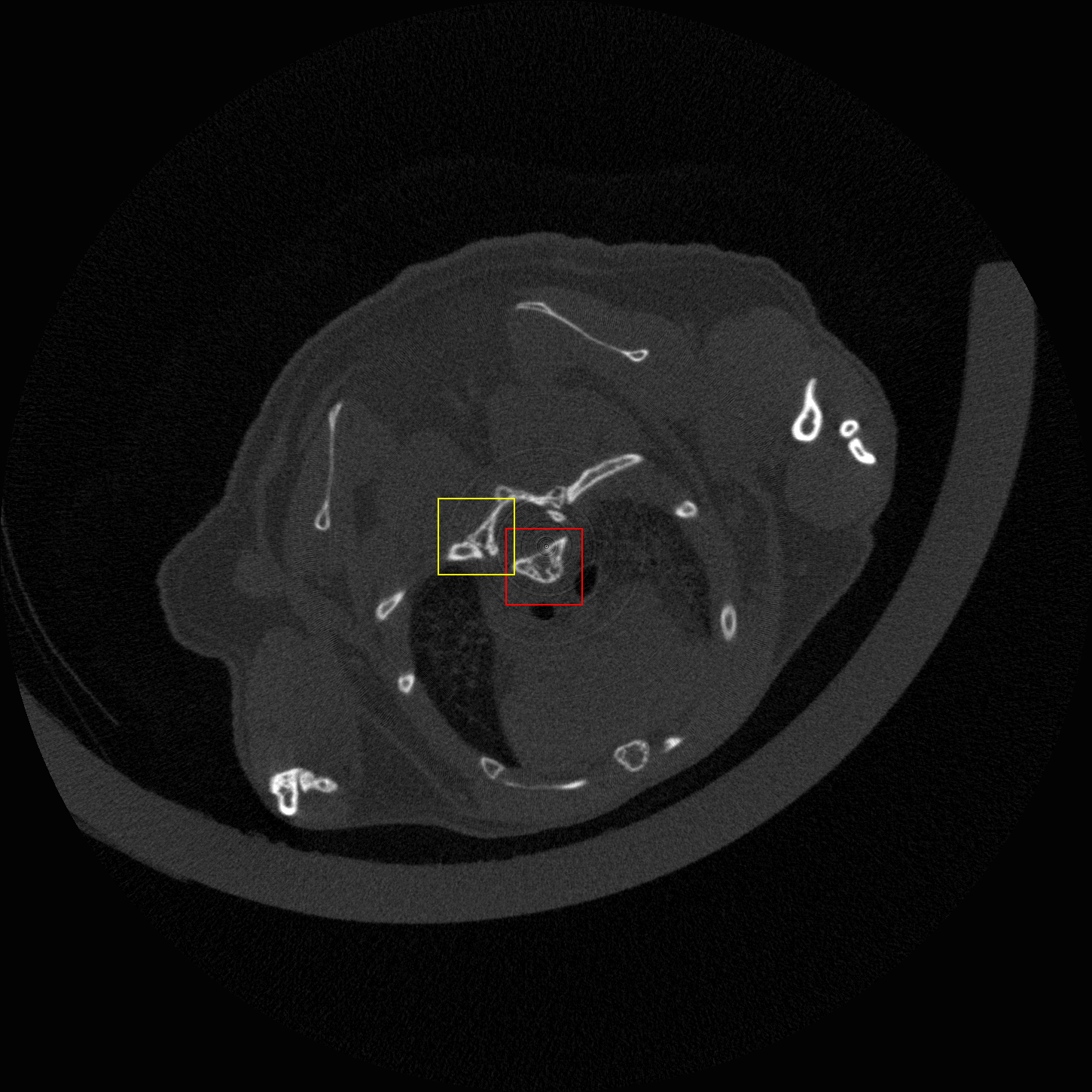}}  ~~
            \subfloat[\scriptsize{Noisy LR}]{\includegraphics[width=1.63in]{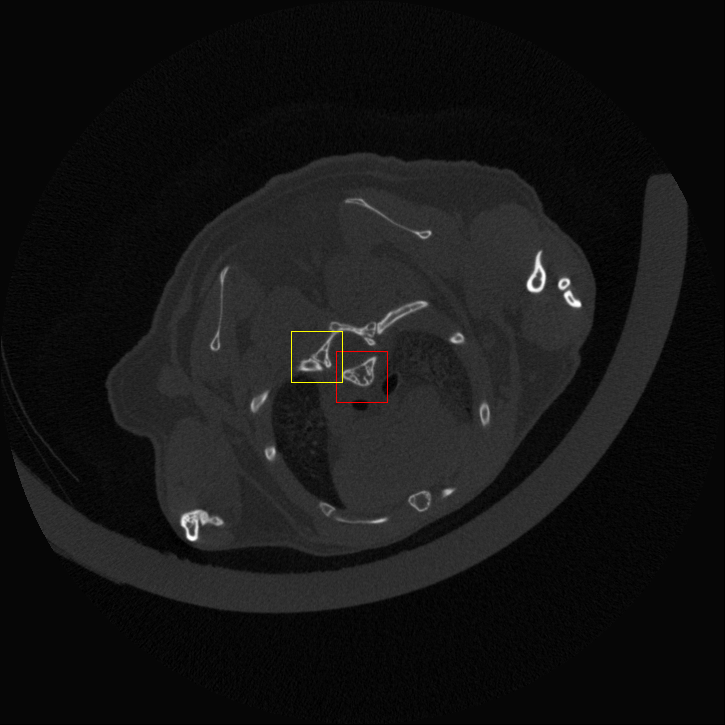}} ~~
            \subfloat[\scriptsize{GAN-CIRCLE$^{s}$}]{\includegraphics[width=1.63in]{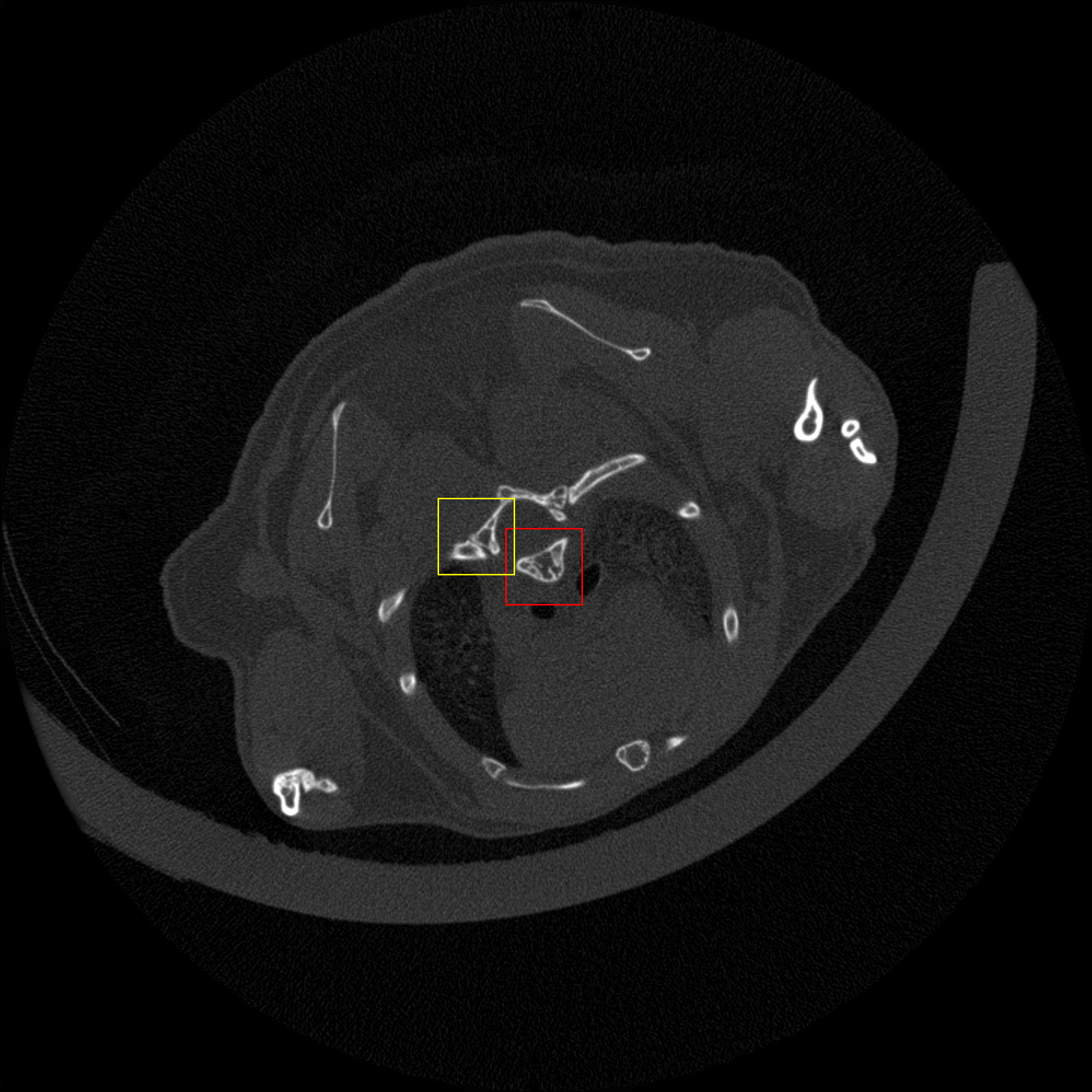}}  ~~
            \subfloat[\scriptsize{GAN-CIRCLE$^{u}$}]{\includegraphics[width=1.63in]{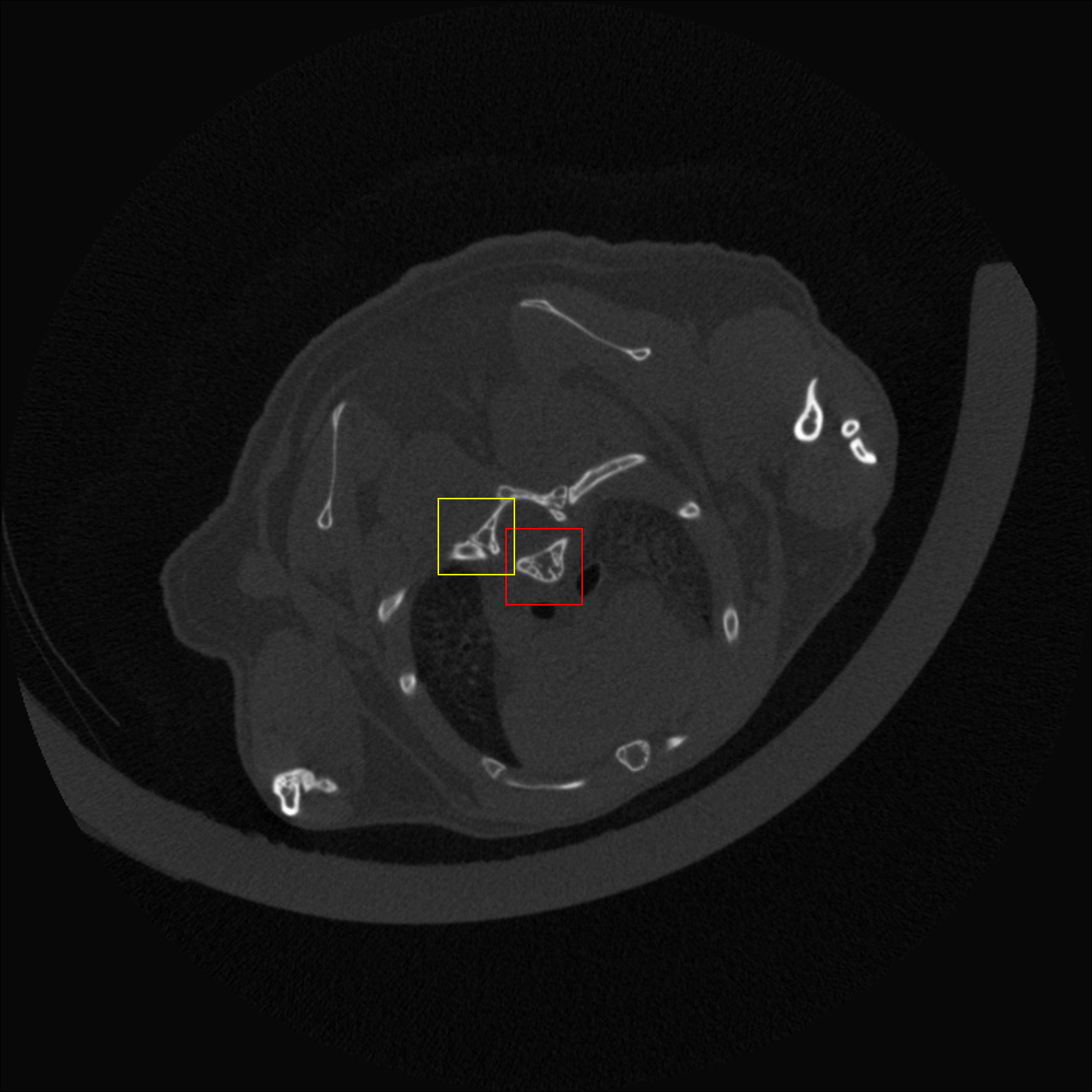}}
         \end{minipage}
         \label{fig: sr_lr6}\\[0.5ex]
        \begin{minipage}[b]{.80\textwidth}
            \centering
            \subfloat[\scriptsize {Noisy LR}]{\includegraphics[width=0.92in]{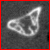}}~
            \subfloat[\scriptsize {NN$^{\pmb{+}}$}]{\includegraphics[width=0.92in]{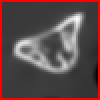}}~
            \subfloat[\scriptsize {Bilinear$^{\pmb{+}}$}]{\includegraphics[width=0.92in]{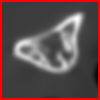}}~
            \subfloat[\scriptsize {Bicubic$^{\pmb{+}}$}]{\includegraphics[width=0.92in]{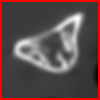}}~
            \subfloat[\scriptsize {Lanczos$^{\pmb{+}}$}]{\includegraphics[width=0.92in]{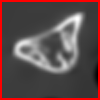}}~
            \subfloat[\scriptsize {A$^{\pmb{+}}$}]{\includegraphics[width=0.92in]{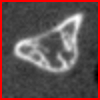}}~
            \subfloat[\scriptsize {FSRCNN}]{\includegraphics[width=0.92in]{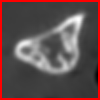}}\\[0.5ex]
            \subfloat[\scriptsize {ESPCN}]{\includegraphics[width=0.92in]{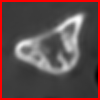}}~
            \subfloat[\scriptsize {LapSRN}]{\includegraphics[width=0.92in]{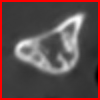}}~
            \subfloat[\scriptsize {SRGAN}]{\includegraphics[width=0.92in]{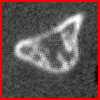}}~
            \subfloat[\scriptsize {G-Fwd}]{\includegraphics[width=0.92in]{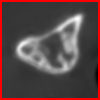}}~
            \subfloat[\scriptsize {G-Adv}]{\includegraphics[width=0.92in]{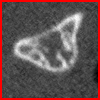}}~
            \subfloat[\scriptsize {\mbox{GAN-CIRCLE$^{s}$}}]{\includegraphics[width=0.92in]{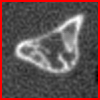}}~
            \subfloat[\scriptsize {\mbox{GAN-CIRCLE$^{u}$}}]{\includegraphics[width=0.92in]{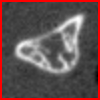}}
        \end{minipage}
        \label{fig: eg6_roi1}\\[0.5ex]
        \begin{minipage}[b]{.80\textwidth}
            \centering
            \subfloat[\scriptsize {Noisy LR}]{\includegraphics[width=0.92in]{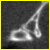}}~
            \subfloat[\scriptsize {NN$^{\pmb{+}}$}]{\includegraphics[width=0.92in]{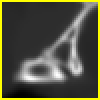}}~
            \subfloat[\scriptsize {Bilinear$^{\pmb{+}}$}]{\includegraphics[width=0.92in]{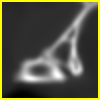}}~
            \subfloat[\scriptsize {Bicubic$^{\pmb{+}}$}]{\includegraphics[width=0.92in]{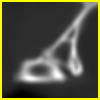}}~
            \subfloat[\scriptsize {Lanczos$^{\pmb{+}}$}]{\includegraphics[width=0.92in]{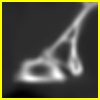}}~
            \subfloat[\scriptsize {A$^{\pmb{+}}$}]{\includegraphics[width=0.92in]{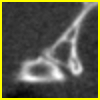}}~
            \subfloat[\scriptsize {FSRCNN}]{\includegraphics[width=0.92in]{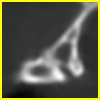}}\\[0.5ex]
            \subfloat[\scriptsize {ESPCN}]{\includegraphics[width=0.92in]{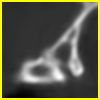}}~
            \subfloat[\scriptsize {LapSRN}]{\includegraphics[width=0.92in]{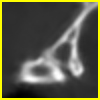}}~
            \subfloat[\scriptsize {SRGAN}]{\includegraphics[width=0.92in]{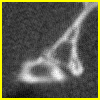}}~
            \subfloat[\scriptsize {G-Fwd}]{\includegraphics[width=0.92in]{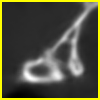}}~
            \subfloat[\scriptsize {G-Adv}]{\includegraphics[width=0.92in]{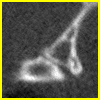}}~
            \subfloat[\scriptsize {\mbox{GAN-CIRCLE$^{s}$}}]{\includegraphics[width=0.92in]{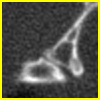}}~
            \subfloat[\scriptsize {\mbox{GAN-CIRCLE$^{u}$}}]{\includegraphics[width=0.92in]{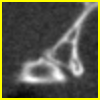}}
        \end{minipage}
        \label{fig: eg6_roi2}
\caption{Visual comparison of SRCT Case~$4$ from the real dataset. The display window is [180, 4096]~HU. The restored bony structures are shown in the red and yellow boxes. (\textbf{Zoomed for visual clarity}).}
\label{fig: example6}
\end{figure*}

\subsubsection{\textbf{Abdominal dataset}}
This clinical dataset is authorized by Mayo Clinic for~\textit{2016 NIH-AAPM-Mayo Clinic Low Dose CT Grand Challenge}. The dataset contains $5,936$ full dose CT images from 10 patients with the reconstruction interval and slice thickness of $0.8~mm$ and $1.0~mm$ respectively. The original CT images were generated by multidetector row CT (MDCT) with image size of $512\times 512$. The projection data is from $2,304$ views per scan. The HR images, with voxel size $0.74\times 0.74\times 0.80~mm^3$, were reconstructed using the FBP method from all $2,304$ projection views. More detailed information of the dataset are given in~\cite{lowdosectgrandchallenge}.

We perform image pre-processing for all CT images through the following workflow. The original CT images were first scaled from the CT Hounsfield Value (HU) to the unit interval [0,1], and treated as the ground-truth HRCT images. In addition, we followed the convention in~\cite{jiang2018super,glasner2009super} to generate LR images by adding noise to the original images and then lowering the spatial resolution by a factor of $2$. For convenience in training our proposed network, we up-sampled the LR image via proximal interpolation to ensure that $\bm{x}$ and $\bm{y}$ are of the same size.

Since the amount of training data plays a significant role in training neural networks~\cite{liu2018learning}, we extracted overlapping patches from LRCT and HRCT images instead of directly feeding the entire CT images to the training pipeline. The overlapped patches were obtained with a predefined sliding size. This strategy preserves local anatomical details, and boost the number of samples. We randomly cropped HRCT images into patches of $64\times 64$, along with their corresponding LRCT patches of size $32\times 32$ at the same center point for supervised learning. With the unsupervised learning methods, the size of the HRCT and LRCT patches are $64\times 64$ in batches of size $64$.

\subsection{Implementation Details}
In the proposed GAN-CIRCLE, we initialized the weights of the Conv layer based on~\cite{he2015delving}. We computed~\textit{std} in the manner of $\sqrt[]{2/m}$ where~\textit{std} is the standard deviation, $m=f_{s}^2\times n_{f}$, $f_{s}$ the filter size, and $n_{f}$ the number of filters.~\textit{i.e.}, given $f_{s} = 3$ and $n_{f} = 16$,~\textit{std} = $0.118$ and all bias were initialized to $0$. In the training process, we empirically set $\lambda_1$, $\lambda_2$, $\lambda_3$ to $1$, $0.5$, $0.001$. Dropout regularization~\cite{hinton2012improving} with $p = 0.8$ was applied to each Conv layer. All the Conv and transposed Conv layers were followed by Leaky ReLu with a slope $\alpha = 0.1$. To make the size of all feature maps the same as that of the input, we padded zeros around the boundaries before the convolution. We utilized the Adam optimizer~\cite{kingma2014adam} with $\beta_1=0.5, \beta_2=0.9$ to minimize the loss function of the proposed network. We set the learning rate to $10^{-4}$ for all layers and then decreased by a factor of $2$ for every $50$ epochs and terminated the training after $100$ epochs. All experiments were conducted using the TensorFlow library on a NVIDA TITAN XP GPU.

\subsection{Performance Comparison}
In this study, we compared the proposed GAN-CIRCLE with the state-of-the-art methods: Nearest-neighbor (NN), Bilinear, Bicubic, Lanczos, adjusted anchored neighborhood regression A$^{\pmb{+}}$~\cite{timofte2014aplus}, FSRCNN~\cite{dong2016accelerating}, ESPCN~\cite{shi2016real}, LapSRN~\cite{LapSRN}, and SRGAN~\cite{ledig2017photo}. For clarity, we categorized the methods into the following classes: the interpolation-based, dictionary-based, PSNR-oriented, and GAN-based methods. Especially, we trained the publicly available FSRCNN, ESPCN, LapSRN and SRGAN with our paired LR and HR images. To demonstrate the effectiveness of the DL-based methods, we first denoised the input LR images and then super-resolved the denoised CT image using the typical interpolation methods: nearest neighbor (NN) up-sampling, bilinear interpolation, bicubic interpolation, lanczos interpolation. BM3D\cite{feruglio2010block} is one of the classic image domain denoising algorithms, which is efficient and powerful. Thus, we preprocessed the noisy LRCT images with BMD3, and then super-solved the denoised images by interpolation methods and A$^{\pmb{+}}$. We refer to interpolation-based methods as NN$^{\pmb{+}}$, Bilinear$^{\pmb{+}}$, Bicubic$^{\pmb{+}}$, Lanczos$^{\pmb{+}}$. 

We evaluated three variations of the proposed method: (1) G-Forward (G-Fwd), which is the forward generator of GAN-CIRCLE, (2) G-Adversarial (G-Adv), which uses the adversarial learning strategy, and (3) the full-fledged GAN-CIRCLE. To emphasize the effectiveness of the GAN-CIRCLE structure, we first trained the three models using the supervised learning strategy, and then trained our proposed GAN-CIRCLE in the semi-supervised scenario (GAN-CIRCLE$^{s}$), and finally implement GAN-CIRCLE in the unsupervised manner (GAN-CIRCLE$^{u}$). In the semi-supervised settings, two datasets were created separately by randomly splitting the dataset into paired and unpaired dataset with respect to three variants: $100\%$, $50\%$, and $0\%$ paired. To better evaluate the performance of each methods, we use the same size of the dataset for training and testing.

We validated the SR performance in terms of three widely-used image quality metrics: Peak signal-to-noise ratio (PSNR), Structural Similarity (SSIM)~\cite{wang2004image}, and Information Fidelity Criterion (IFC)~\cite{sheikh2005information}. Through extensive experiments, we compared all the above-mentioned methods on the two benchmark datasets described in Section~\ref{subsec:dataset}.

\subsection{Experimental Results with the Tibia Dataset}
\label{subsec:Tibiadataset}
We evaluated the proposed algorithms against the state-of-the-art algorithms on the tibia dataset. We present typical results in Fig.~\ref{fig: example1}. It is observed that BM3D can effectively remove the noise, but it over-smoothens the noisy LR images. Then, the interpolation-based methods (NN$^{\pmb{+}}$, Bilinear$^{\pmb{+}}$, Bicubic$^{\pmb{+}}$, Lanczos$^{\pmb{+}}$) yield noticeable artifacts caused by partial aliasing. On the other hand, the DL-based methods suppress such artifacts effectively. It can be seen that our proposed GAN-CIRCLE recovers more fine subtle details and captures more anatomical information in Fig.~\ref{fig: example1_roi}. It is worth mentioning that Fig.~\ref{fig: example1} shows that there are severe distortions of the original images but SRGAN generates compelling results in Figs.~\ref{fig: example2}-\ref{fig: example6}, which indicate VGG network is a task-specific network which can generate images with excellent image quality. We argue that the possible reason is that the VGG network~\cite{simonyan2014very} is a pre-trained CNN-based network based on natural images with structural characteristic correlated with the content of medical images~\cite{shen2017deep}. Fig.~\ref{fig: example1_roi} presents that the proposed GAN-CIRCLE$^{s}$ can predict images with shaper boundaries and richer textures than GAN-CIRCLE, and GAN-CIRCLE$^{u}$ which learns additional anatomical information from the unpaired samples. The quantitative results are in Table~\ref{table: psnr&ssim&ifc}. The results demonstrate that the G-Forward achieves the highest scores using the evaluation metrics, PSNR and SSIM, which outperforms all other methods. However, it has been pointed out in~\cite{yang2018low,shan20183d} that high PSNR and SSIM values cannot guarantee a visually favorable result. Non-GAN based methods (FSRCNN, ESPCN, LapSRN) may fail to recover some fine structure for diagnostic evaluation, such as shown by zoomed boxes in Fig.~\ref{fig: example1_roi}. Quantitatively, GAN-CIRCLE achieves the second best values in terms of SSIM and IFC. It has been pointed out in~\cite{yang2014single} that IFC value is correlated well with the human perception of SR images. Our GAN-CIRCLE$^{s}$ obtained comparable results qualitatively and quantitatively. Table~\ref{table: psnr&ssim&ifc} shows that~\textbf{the proposed semi-supervised method performs similarly compared to the fully supervised methods on the tibia dataset}. In general, our proposed GAN-CIRCLE can generate more pleasant results with sharper image contents.

\begin{table*}[!t]
\renewcommand{\arraystretch}{0.8}
\centering
\caption{Diagnostic quality assessment in terms of subjective quality scores for different algorithms (mean$\pm$stds).~\textbf{{\color{red} Red}} and {\color{blue} {blue}} indicate the best and the second best performance, respectively.}
 \setlength{\tabcolsep}{0.8mm}
\begin{tabular}{c c c c c c c c c c c c}
\hline\hline
& \multicolumn{5}{c}{Tibia Dataset} && \multicolumn{5}{c}{Abdominal Dataset} \\
&~\tiny{\textbf{Image Sharpness}} &~\tiny{\textbf{Image Noise}} &\tiny{\textbf{Contrast Resolution}} &~\tiny{\textbf{Diagnostic Acceptance}} &~\tiny{\textbf{Overall Quality}} &&~\tiny{\textbf{Image Sharpness}} &~\tiny{\textbf{Image Noise}} &\tiny{\textbf{Contrast Resolution}} &~\tiny{\textbf{Diagnostic Acceptance}} &~\tiny{\textbf{Overall Quality}}  \\
\cline{2-6}\cline{8-12}
\scriptsize{NN$^{\pmb{+}}$} 		& 1.89$\pm$0.27 			& 2.43$\pm$0.21 	& 1.89$\pm$0.78 			& 1.52$\pm$0.69 			& 1.72$\pm$0.33 
			&& 1.98$\pm$0.46 			& 3.22$\pm$1.45 	& 1.47$\pm$0.23 			& 1.67$\pm$0.85 			& 2.34$\pm$0.42\\
\scriptsize{Bilinear$^{\pmb{+}}$} 		& 1.87$\pm$0.41 			& 2.52$\pm$0.73 	& 2.01$\pm$0.83 			& 1.65$\pm$0.73 			& 1.95$\pm$0.47 
			&& 2.02$\pm$0.41 			& 3.12$\pm$1.58 	& 1.85$\pm$0.96 			& 1.75$\pm$0.83 			& 2.43$\pm$0.45\\
\scriptsize{Bicubic$^{\pmb{+}}$} 		& 2.55$\pm$0.43 			& 2.34$\pm$0.82 	& 2.19$\pm$0.91 			& 1.82$\pm$0.21 			& 2.12$\pm$0.23 
			&& 2.52$\pm$0.53 			& 2.87$\pm$1.05 	& 2.51$\pm$0.53 			& 2.27$\pm$0.45 			& 2.61$\pm$0.67\\
\scriptsize{Lanczos$^{\pmb{+}}$} 	& 2.25$\pm$0.39 			& 2.50$\pm$0.85 	& 2.36$\pm$0.82 			& 1.93$\pm$0.43 			& 2.23$\pm$0.29
			&& 2.53$\pm$0.59 			& 2.99$\pm$0.86 	& 2.55$\pm$0.64 			& 2.21$\pm$0.35 			& 2.68$\pm$0.31\\
\scriptsize{A$^{\pmb{+}}$} 	& 2.34$\pm$0.47 			& 2.54$\pm$0.68 	& 2.52$\pm$0.67 			& 1.98$\pm$0.59 			& 2.37$\pm$0.97
			&& 2.74$\pm$0.75 			& 3.07$\pm$0.96 	& 2.61$\pm$0.69 			& 2.35$\pm$0.57 			& 2.74$\pm$0.71\\
\scriptsize{FSRCNN} 		& 2.85$\pm$0.94 			& 3.16$\pm$0.57 	& 2.54$\pm$0.96 			& 2.77$\pm$0.69 			& 3.27$\pm$0.76
			&& 3.07$\pm$0.89 			& {\color{blue}{3.55$\pm$0.50}} 	& 2.94$\pm$0.78 			& 2.92$\pm$0.58 			& 3.09$\pm$0.53\\
\scriptsize{ESPCN} 		& 2.82$\pm$0.86 			& 3.18$\pm$0.51 	& 2.58$\pm$0.46 			& 2.95$\pm$0.46 			& 3.49$\pm$0.66
			&& 2.95$\pm$1.43 			& 3.39$\pm$0.80 	& 2.85$\pm$0.63 			& 2.76$\pm$0.83 			& 3.06$\pm$0.85\\
\scriptsize{LapSRN} 		&  2.91$\pm$0.88 			& {\color{blue}{3.49$\bm{\pm}$0.70}} 	& 2.69$\pm$0.56 			& 3.01$\pm$0.78 			& 3.63$\pm$0.61
			&& 3.01$\pm$0.56 			& \textbf{{\color{red}3.58$\bm{\pm}$0.81}} 	& 2.83$\pm$0.71 			& 3.25$\pm$0.92 			& 3.11$\pm$0.78\\
\scriptsize{SRGAN}	 	&  1.94$\pm$0.37			& 2.71$\pm$0.23 	& 1.91$\pm$0.71 			& 1.75$\pm$0.83			& 1.93$\pm$1.01
			&& 3.35$\pm$0.97 			& 3.23$\pm$1.01 	& 3.27$\pm$0.92 			& 3.46$\pm$1.11 			& 3.41$\pm$0.94\\
\scriptsize{G-Fwd} 		& 2.99$\pm$0.42 			& \textbf{{\color{red}3.59$\bm{\pm}$0.57}} 	& 3.07$\pm$0.91 			& 3.45$\pm$1.02			& 3.70$\pm$0.71
			&& 3.25$\pm$0.94 			& 3.53$\pm$0.70 	& 2.95$\pm$0.57 			& 3.38$\pm$0.93 			& 3.09$\pm$0.55\\
\scriptsize{G-Adv}	 	& 2.89$\pm$0.86			& 3.13$\pm$1.02	& 3.02$\pm$0.58			& 3.29$\pm$0.69			& 3.62$\pm$0.67
			&& 3.45$\pm$1.12 			& 3.34$\pm$0.81 	& 3.31$\pm$0.86 			& 3.48$\pm$0.77 			& 3.32$\pm$0.82\\
\scriptsize{GAN-CIRCLE} 	& \textbf{{\color{red}3.12$\bm{\pm}$0.73}}	& 3.40$\pm$0.43	&  \textbf{{\color{red}3.17$\bm{\pm}$0.46}}	& \textbf{{\color{red}3.61$\bm{\pm}$0.36}}	& \textbf{{\color{red}3.79$\bm{\pm}$0.72}} 
			&& \textbf{{\color{red}3.59$\bm{\pm}$0.41}} 			& 3.41$\pm$0.42 	& \textbf{{\color{red}3.51$\bm{\pm}$0.66}} 			& \textbf{{\color{red}3.64$\bm{\pm}$0.54}} 			& \textbf{{\color{red}3.62$\bm{\pm}$0.41}}\\

\scriptsize{GAN-CIRCLE$^{s}$} 	& {\color{blue}{3.02$\pm$0.78}}	& 3.14$\pm$0.68	& {\color{blue}{3.12$\pm$0.88}}	& {\color{blue}{3.47$\pm$0.67}}	& {\color{blue}{3.71$\pm$0.76}}
			&& {\color{blue}{3.48$\pm$0.81}} 			& 3.29$\pm$0.80 	& {\color{blue}{3.42$\pm$0.78}} 			& {\color{blue}{3.57$\pm$0.68}} 			& {\color{blue}{3.51$\pm$0.46}}\\
\scriptsize{GAN-CIRCLE$^{u}$}	 	& 2.91$\pm$0.82			& 3.32$\pm$0.89	& 3.08$\pm$0.94			& 3.32$\pm$0.48			& 3.57$\pm$0.52
			&& 3.46$\pm$0.73 			& 3.39$\pm$1.04 	& 3.39$\pm$0.50 			& 3.54$\pm$0.53 			& 3.34$\pm$1.01\\
\hline\hline
\end{tabular}
\label{table: reader}
\end{table*}

\subsection{Experimental Results on the Abdominal Dataset}
\label{subsec:Abdominaldataset}
We further compared the above-mentioned algorithms on the abdominal benchmark dataset. A similar trend can be observed on this dataset. Our proposed GAN-CIRCLE can preserve better anatomical informations and more clearly visualize the portal vein as shown in Fig.~\ref{fig: example2}. These results demonstrate that PSNR-oriented methods (FSRCNN, ESPCN, LapSRN) can significantly suppress the noise and artifacts. However, it suffers from low image quality as judged by the human observer since it assumes that the impact of noise is independent of local image features, while the sensitivity of the Human Visual System (HVS) to noise depends on local contrast, intensity and structural variations. Fig.~\ref{fig: example2} displays the LRCT images processed by GAN-based methods (SRGAN, G-Adv, GAN-CIRCLE, GAN-CIRCLE$^{s}$, and GAN-CIRCLE$^{u}$) with improved structural identification. It can also observed that the GAN-based models also introduce strong noise into results. For example, there exist tiny artifacts on the results of GAN-CIRCLE$^{u}$. As the SR results shown in Fig.~\ref{fig: example2}, our proposed approaches (GAN-CIRCLE, GAN-CIRCLE$^{s}$) are capable of retaining high-frequency details to reconstruct more realistic images with relatively lower noise compared with the other GAN-based methods (G-Adv, SRGAN). Table~\ref{table: psnr&ssim&ifc} show that G-Fwd achieves the best performance in PSNR. Our proposed methods GAN-CIRCLE and GAN-CIRCLE$^{s}$ both obtain the pleasing results in terms of SSIM and IFC. In other words, the results show that the proposed GAN-CIRCLE and GAN-CIRCLE$^{s}$ generate more visually pleasant results with sharper edges on the abdominal dataset than the competing state-of-the-art methods. 

\subsection{Super-resolving Real-world Images}
\label{subsec:microanimialdataset}
We analyzed the performance of the SR methods in the simulated SRCT scenarios in Sections~\ref{subsec:Tibiadataset} and~\ref{subsec:Abdominaldataset}. These experimental results show that the DL-based methods are very effective in addressing the ill-posed SRCT problems with two significant features.~\textit{First}, SRCT aims at recovering a HRCT image from a LRCT images under a low-dose protocol.~\textit{Second}, most DL-based methods assume the paired LRCT images and HRCT images are matched, an assumption which is likely to be violated in clinical practice. In other words, the above-evaluated datasets were simulated, and thus the fully supervised algorithms can easily cope with SRCT tasks, with~\textbf{exactly matched} training samples. Our further goal is to derive the~\textbf{semi-supervised} scheme to handle unmatched/unpaired data with a relative lack of matched/paired data to address real SRCT tasks. In this subsection, we demonstrate a strong capability of the proposed methods in the real applications using~\textbf{a small amount of mismatched} paired LRCT and HRCT images as well as a high flexibility of adapting to various noise distributions.
\subsubsection{\textit{Practical SRCT Implementation Details}} 
We first obtained LRCT and HRCT images using a deceased mouse on the same scanner with two scanning protocols. The micro-CT parameters are as follows: X-ray source circular scanning, $60$ kVp, $134$ mAs, $720$ projections over a range of $360$ degrees, exposure $50$ ms per projection, and the micro-CT images were reconstructed using a conventional filtered back projection algorithm (FDK): HRCT image size $1450\times 1450$, $600$ slices at $48~\mu m$ isotropic voxel size, and the LRCT image size $725\times 725$, $300$ slices at $24~\mu m$ isotropic voxel size. Then, we compared with the state-of-the-art super-resolution methods. Since the real data are unmatched, we accordingly evaluated our proposed GAN-CIRCLE$^{s}$ and GAN-CIRCLE$^{u}$ networks for $1$X resolution improvement.
\subsubsection{\textit{Comparison With the State-of-the-Art methods}} 
The quantitative results were summarized for all the involved methods in Table~\ref{table: psnr&ssim&ifc}. The PSNR-oriented approaches, such as FSRCNN, ESPCN, LapSRN, and our G-Fwd, yield higher PSNR and SSIM values than the GAN-based methods. It is not surprising that the PSNR-oriented methods obtained favorable PSNR values since their goal is to minimize per-pixel distance to the ground truth. However, our GAN-CIRCLE$^{s}$ and GAN-CIRCLE$^{u}$ achieved the highest IFC among all the SR methods. Our method GAN-CIRCLE$^{s}$ obtained the second best results in term of SSIM. The visual comparisons are given in Figs.~\ref{fig: example5} and~\ref{fig: example6}. To demonstrate the robustness of our methods, we examined anatomical features in the lung regions and the bone structures of the mice, as shown in Figs.~\ref{fig: example5} and~\ref{fig: example6} respectively. It is observed that the GAN-based approaches performed favorably over the PSNR-oriented methods in term of perceptual quality as illustrated in Figs.~\ref{fig: example5} and~\ref{fig: example6}. Fig.~\ref{fig: example5} confirms that the PSNR-oriented methods produced blurry results especially in the lung regions, while the GAN-based methods restored anatomical contents satisfactorily. In Fig.~\ref{fig: example6}, it is notable that our methods GAN-CIRCLE$^{s}$ and GAN-CIRCLE$^{u}$ performed better than the other methods in terms of recovering structural information and preserving edges. These SR results demonstrate that our proposed methods can provide better visualization of bone and lung microarchitecture with sharp edge and rich texture.

\subsection{Diagnostic Quality Assessment}
We invited three board-certified radiologists with mean clinical CT experience of 12.3 years to perform independent qualitative image analysis on $10$ sets of images from two benchmark dataset (Tibia and Abdominal Dataset). Each set includes the same image slice but generated using different methods. We label HRCT and LRCT images in each set as reference. The $10$ sets of images from two datasets were randomized and deidentified so that the radiologists were blind to the post-proprocessing algorithms. Image sharpness, image noise, contrast resolution, diagnostic acceptability and overall image quality were graded on a scale from 1 (worst) to 5 (best). A score of 1 refers to a `non-diagnostic' image, while a score of 5 means an `excellent' diagnostic image quality. The mean scores with their standard deviation are presented in Table~\ref{table: reader}. The radiologists confirmed that~\textbf{GAN-based} methods (G-Adv, SRGAN, GAN-CIRCLE, GAN-CIRCLE$^{s}$ and GAN-CIRCLE$^{u}$) provide sharper images with better texture details, while~\textbf{PSNR-oriented} algorithms (FSRCNN, ESPCN, LapSRN, G-Fwd) give the higher noise suppression scores. Table~\ref{table: reader} shows that our proposed GAN-CIRCLE and GAN-CIRCLE$^{s}$ achieve comparable results, while outperforming the other methods in terms of image sharpness, contrast resolution, diagnostic acceptability and overall image quality.

\section{Discussions}
\label{sec:discussion}
SR imaging holds tremendous promise for practical medical applications;~\textit{i.e.}, depicting bony details, lung structures, and implanted stents, and potentially enhancing radiomics analysis. As a results, X-ray computed tomography can provide compelling practical benefit in biological evaluation.

High resolution micro-CT is well-suited for bone imaging. Osteoporosis, characterized by reduced bone density and structural degeneration of bone, greatly diminishes bone strength and increases the risk of fracture~\cite{cummings2002epidemiology}. Histologic studies have convincingly demonstrated that bone micro-structural properties are strong determinants of bone strength and fracture risk~\cite{kleerekoper1985role,legrand2000trabecular,parfitt1983relationships}. Modern whole-body CT technologies, benefitted with high spatial resolution, ultra-high speed scanning, relatively-low dose radiation, and large scan length, allows quantitative characterization of bone micro-structure~\cite{chen2018quantitative}. However, the state-of-the-art CT imaging technologies only allow the spatial resolution comparable or slightly higher than human trabecular bone thickness ($100-200~\mu m^6$~\cite{ding2000quantification}) leading to fuzzy representation of individual trabecular bone micro-structure with significant partial volume effects that adds significant errors in measurements and interpretations. The spatial resolution improvements in bone micro-structural representation will largely reduce such errors and improve generalizability of bone micro-structural measures from multi-vendor CT scanners by homogenizing spatial resolution. 

Besides revealing micro-architecture, CT scans of the abdomen and pelvis are diagnostic imaging tests used to help detect diseases of the small bowel and colon, kidney stone, and other internal organs, and are often used to determine the cause of unexplained symptoms. With rising concerns over increased lifetime risk of cancer by radiation dose associated with CT, several studies have assessed manipulation of scanning parameters and newer technologic developments as well as adoption of advanced reconstruction techniques for radiation dose reduction~\cite{kalra2004strategies,marin2009low,prakash2010reducing,vasilescu2012assessment,iyer2016quantitative}. However, in practice, the physical constraints of system hardware components and radiation dose considerations constrain the imaging performance, and computational means are necessary to optimize image resolution. For the same reason, high-quality/high-dose CT images are not often available, which means that there are often not enough paired data to train a hierarchical deep generative model. 

Our results have suggested an interesting topic on how to utilize unpaired data so that the imaging performance could be improved. In this regard, the use of the adversarial learning as the regularization term for SR imaging is a new mechanism to capture anatomical information. However, it should be noted that the existing GAN-based methods introduce additional noise to the results, as seen in Section~\ref{subsec:Tibiadataset} and~\ref{subsec:Abdominaldataset}. To cope with this limitation, we have incorporated the cycle-consistency so that the network can learn a complex deterministic mapping to improve image quality. The enforcement of identity and supervision allows the model to master more latent structural information to improve image resolution. Also, we have used the Wasserstein distance to stabilize the GAN training process. Moreover, typical prior studies used complex inference to learn a hierarchy of latent variables for HR imaging, which is hard to be utilized in medical applications. Thus, we have designed an efficient CNN-based network with skip-connection and network in network techniques. In the feature extraction network, we have optimized the network structures and reduced the computational complexity by applying a small amount of filters in each Conv layer and utilizing the ensemble learning model. Both local and global features are cascaded through skip connections before being fed into the restoration/reconstruction network.

Although our model has achieved compelling results, there still exist some limitations. First, the proposed GAN-CIRCLE requires much longer training time than other standard GAN-based methods, which generally requires 1-2 days. Future work in this aspect should consider more principled ways of designing more efficient architectures that allow for learning more complex structural features with less complex networks at less computational cost and lower model complexity. Second, although our proposed model can generate more plausible details and better  anatomical details, all subtle structures may not be always faithfully recovered. It has been also observed that the recent literature~\cite{bellemare2017cramer} mentions that the Wasserstein distance may yield the biased sample gradients, is subject to the risk of incorrect minimum, and not well suitable for stochastic gradient descent searching. In the future, experimenting with the variants of GANs are highly recommended. Finally, we notice that the network with the adversarial training can generate more realistic images. However, the restored images cannot be uniformly consistent to the original high-resolution images. To make further progress, we may also undertake efforts to add more constraints such as the sinogram consistence and the low-dimensional manifold constraint to decipher the relationship between noise, blurry appearances of images and the ground truth, and even develop an adaptive and/or task-specific loss function.

\section{Conclusions}
\label{sec:conclusions}
In this paper, we have estabilished a cycle wasserstein regression adversarial training framework for CT SR imaging. Aided by unpaired data, our approach learns complex structured features more effectively with a limited amount of paired data. At a low computational cost, the proposed network G-Forward can achieve the significant SR gain. In general, the proposed GAN-CIRCLE has produced promising results in terms of preserving anatomical information and suppressing image noise in a purely supervised and semi-supervised learning fashion. Visual evaluations by the expert radiologists confirm that our proposed GAN-CIRCLE networks have brought superior diagnostic quality, which is consistent to systematic quantitative evaluations in terms of traditional image quality measures. 

\section*{Acknowledgment}
The authors would like to thank the NVIDIA Corporation for the donation of the TITAN XP GPU to Dr. Ge Wang's laboratory, which was used for this study. The authors would like to thank Dr. Shouhua Luo (Southeast University, China) for providing small animal data collected on an~\textit{in vivo} micro-CT system.

\bibliographystyle{IEEEtran}
\bibliography{srct}
\ifCLASSOPTIONcaptionsoff
  \newpage
\fi

\end{document}